\documentclass[12pt]{article}

\usepackage[utf8]{inputenc} 
\usepackage[hidelinks]{hyperref}       
\usepackage{url}            
\usepackage{booktabs}       
\usepackage{amsfonts}       
\usepackage{nicefrac}       
\usepackage{microtype}      
\usepackage{lipsum}
\usepackage{graphicx}
\usepackage{algorithm}
\usepackage{algorithmic}
\usepackage{amsmath}
\usepackage{amssymb}
\usepackage[margin=1in]{geometry}
\usepackage{siunitx}
\usepackage{threeparttable}
\usepackage{caption}
\usepackage{multirow}
\usepackage{subcaption}
\usepackage[shortlabels]{enumitem}
\usepackage{newtxtext}
\usepackage{bm}
\usepackage{pgfplots}
\usepackage{xr-hyper}
\usepackage{natbib}
\pgfplotsset{compat=1.17}
\graphicspath{ {./images/} }
\usepackage{setspace}
\newtheorem{theorem}{Theorem}
\newtheorem{lemma}{Lemma}
\newtheorem{rem}{Remark}

\title{Observation-Level Watermarking and Detection for Tabular Data}

\author{
  Dongyu Cui\thanks{School of Statistics, University of Minnesota}
  \and
  Xuan Bi\thanks{Carlson School of Management, University of Minnesota, email: \texttt{xbi@umn.edu}}
}
\date{}

\begin{document}
\doublespacing
\maketitle

\begin{abstract}
\noindent With the development of generative AI, watermarking techniques have been widely used to detect the authenticity of AI-generated data and protect the rights of users and creators. While it is already well applied in data types including imaging and text data, watermarking tabular data is still under-explored. Existing methods primarily focus on numerical data, leaving discrete, categorical, and mixed data less studied. In this work, we propose STAMP (Single-observation Tabular Attribution and Marking Procedure), a novel framework for watermarking tabular data that can accommodate and preserve a wide range of distributions. We also develop a corresponding detection mechanism, which can reliably identify watermarks even when the sample size is as small as one. We establish theoretical guarantees for asymptotic consistency and detection accuracy. Finally, through extensive simulation studies and two real-data applications, we demonstrate that the proposed method is effective and robust to subsetting, while maintaining data fidelity and a high detection rate.
\end{abstract}


\section{Introduction} \label{sec:introduction}

In modern society, data are produced and shared at an unprecedented rate. With the rapid development of generative AI techniques, it has become increasingly easy to generate various types of data, including images, text, and tabular data. This development has raised several important issues. The first issue is about data authenticity. On the one hand, AI-generated data can be highly realistic and easily mistaken for real data. On the other hand, datasets may be provided without a clear source and misrepresented as originating from reputable individuals or institutions. Both cases highlight the need for reliable authenticity verification.
Second, data may be used by others without the creator’s permission, raising concerns about ownership and unauthorized use. 
Third, even when data are shared, creators may want to seek proper attribution, as malicious recipients may redistribute the data while falsely claiming it as their own. 
Last, ensuring traceability is also important, so that the origin of a dataset can be identified and responsible parties can be held accountable for misuse or leakage.
These needs call for techniques to make data identifiable, and thereby protect the rights of users and creators.

Watermarking is a widely used technique to address these issues. It involves embedding a unique identifier into the data, which later can be used to verify the data's origin. 
Specifically, a watermarking method usually consists of two steps: \emph{watermark insertion} and \emph{watermark detection}. First, a unique identifier (i.e., a \emph{key}) is inserted into the data to make it identifiable. 
Second, a detection procedure is used to determine the presence of the key.

In recent years, watermarking methods have been proposed for different data types and modalities. For image data, 
commonly used methods include frequency-domain methods (e.g., signal encryption after Fourier transformation) \citep{altun2009optimal,hernandez2000dct} and recent neural-network-based methods \citep{gunn2025undetectable, Tancik_2020_CVPR, xian2024raw,Zhu_2018_ECCV}. 
For text data, watermarking techniques 
include synonym substitution \citep{yang2023watermarking}, sentence reordering \citep{atallah2001natural}, the green-red list \citep{kirchenbauer2023watermark}, or character-level modifications \citep{por2012unispach}. Specifically, for watermarking text generated by large language models, several methods have been proposed, such as modifying the token selection process during text generation or introducing specific patterns in the generated text \citep{WatermarkingLargeLanguage2023,hu2024unbiased,li2025statistical}. These methods aim to make the watermark robust against paraphrasing and other text transformations while preserving the readability and coherence of the text.
On the other hand, designing and detecting watermarks may encounter several challenges. 
First, a watermarking method usually entails a tradeoff between detectability and utility \citep{zizhuo2024robust}. That is, a strong watermark (e.g., a visible image watermark) can be easily detected,
but may inadvertently alter the original data and hence reduce its utility. 
Second, a successful watermark should also be robust against modifications. 
Such modifications include watermark removal \citep{
tian2024perceptive}, where adversaries intend to 
remove an embedded watermark,
and watermark forgery, 
where adversaries attempt to forge watermarks that falsely attribute ownership \citep{
muller2025black}.


Despite the significant progress in watermarking techniques for imaging and text data, research on watermarking tabular data remains limited. Yet, tabular data play a critical role in numerous applications, including healthcare, finance, and the social sciences, where ensuring data authenticity is essential.
Several tabular data watermarking methods have been proposed. For example, He et al. \citep{he2024watermarking} and Daniel et al. \citep{ngo2024adaptive} adopt the idea of green-red lists from text data, which watermark numerical data and modify the values of data points only slightly. Zheng et al. \citep{zheng2024tabularmark} propose a new scheme named TabularMark, which randomly selects some entries to add noise while preserving data utility.
In addition, several deep-learning-based watermarking methods have also been proposed. TAB-DRW \citep{zhao2025tab} embeds watermark via discrete Fourier transformation. TabWak \citep{zhu2025tabwak} achieves watermark embedding by adding the key into a diffusion model to generate tabular data, and develop an inverse model to realize detection. MUSE \citep{fang2025muse} extends TabWak by generating multiple candidate datasets, and selects the one that improves robustness while preserving detectability.

However, several critical challenges remain in watermarking tabular data. First, to the best of our knowledge, there is no unified watermarking framework for both univariate and multivariate data, as well as for both continuous and discrete variables.
Second, for extremely small samples (e.g., those that only contain a few or a few dozen instances), 
there are still no effective detection methods for identifying embedded watermarks.
Third, most existing watermarks operate at the dataset level, making it difficult to attribute individual data instances to specific users.

This article proposes STAMP (Single-observation Tabular Attribution and Marking Procedure), a novel watermark insertion and detection framework for tabular data. 
Inspired by the distribution-invariant transformation method \citep{bi2023distribution}, we design an insertion method that can effectively embed watermarks into various types of data while preserving their underlying distributions. More importantly, we also develop a novel detection method, which can effectively detect the inserted watermarks.
Specifically, the proposed insertion method is based on data transformation that involves the empirical cumulative distribution function and key injection. Accordingly, the detection is achieved via reverse transformation and key matching.
Theoretically, we show that the detection rate converges to 1 asymptotically, and establish asymptotic consistency, in the sense that the watermarked data preserve the original distribution.
Numerically, we evaluate our method on multiple simulated and real datasets and demonstrate its effectiveness in terms of fidelity preservation, detection accuracy, and robustness against subsetting.

The proposed method has several key advantages. First, it provides a unified framework that accommodates univariate and multivariate data, as well as both continuous and discrete variables. Second, STAMP operates at the level of individual observations for both watermark insertion and detection. This allows reliable detection even for extremely small samples. In the most extreme case, the proposed detection method can determine whether a single value contains the proposed watermark. Third, this observation-level design can be generalized to support user attribution. That is, we can not only detect the presence of watermarks, but also link watermarked data to specific individuals. Finally, the proposed approach preserves the original data distribution asymptotically, ensuring high data fidelity for downstream use.


The rest of the article is organized as follows. In Section \ref{sec:setup and background}, we introduce the problem setup and relevant background information. In Section \ref{sec:method}, we present our proposed watermark insertion and detection methods. Section \ref{sec:theory} studies the theoretical properties of our method. In Section \ref{sec:simulation}, we evaluate our method on multiple simulated datasets. In Section \ref{sec:real_data}, we apply our method to two real datasets and demonstrate its effectiveness. Finally, we conclude the paper in Section \ref{sec:discussion}.

\section{Problem Setup and Background} \label{sec:setup and background}
\subsection{Problem Setup} \label{sec:setup}

In this subsection, we formally formulate the watermarking problem and introduce the relevant notation.
We consider a raw dataset $\mathbf{X}_{n \times p}$, which will be watermarked. Suppose each row of $\mathbf{X}$ is $\mathbf{X}_i=(X_{i1},X_{i2},\ldots,X_{ip})$ that satisfies $\mathbf{X}_i \stackrel{iid}{\sim} F$ for $i=1,\ldots,n$, where $p$ is the number of variables and $F$ is the $p$-dimensional cumulative distribution function (CDF). 
Here each of the $p$ variables can be either continuous or discrete. A categorical variable of $d$ categories should be converted into $d-1$ binary dummy variables before being included in $\mathbf{X}$.

Typically, a watermarking method consists of two steps. The first step is \emph{watermark insertion}. In this step, the goal is to embed a watermark $\mathbf{e}$ into the dataset $\mathbf{X}$, resulting in a watermarked dataset $\tilde{\mathbf{X}}$ of the same size $n \times p$. In other words, we want to design an insertion function $W:\mathbf{X}\rightarrow \tilde{\mathbf{X}}$, such that
$$
\tilde{\mathbf{X}}=W(\mathbf{X},\mathbf{e}),
$$
where $\mathbf{e}$, also known as the \emph{key}, is a series of values. For example, in the proposed method (i.e., Section \ref{sec:method}), $\mathbf{e}$ is a random matrix of the same size $n \times p$. In general, $\mathbf{e}$ serves as the signal to be detected and should remain confidential, as revealing it would compromise the security of the watermark. 
Ideally, the watermarked dataset $\tilde{\mathbf{X}}$ should satisfy fidelity. That is, the distribution of $\tilde{\mathbf{X}}$ should not deviate, or deviate only minimally, from the original distribution $F$. This guarantees that the statistical properties of the original data are approximately preserved and the watermarked dataset retains similar usefulness for any downstream statistical or machine-learning tasks. Additionally, in some applications, it is also desirable for the correlation between $\mathbf{X}_i$ and $\tilde{\mathbf{X}}_i$ to remain high for each $i=1,\ldots,n$,
which enables future data integration or linkage, should additional features for each $i$ become available.

The second step is \emph{watermark detection}. This step intends to design a detection function $D$, such that for any dataset $\mathbf{Z}$, it can detect if the watermark $\mathbf{e}$ is embedded. For example, a detection function should satisfy $D(\tilde{\mathbf{X}},\mathbf{e})=1$ and $D(\mathbf{X},\mathbf{e})=0$. Generally, one could write $D$ as 
\begin{equation} \label{eq:detection_general}
    D(\mathbf{Z},\mathbf{e})=\begin{cases}
1 \quad (\text{watermarked}), & \text{if } (\mathbf{Z},\mathbf{e})\in C \\
0 \quad (\text{unwatermarked}), & \text{otherwise,}
\end{cases}
\end{equation}
where $C$ is an acceptance region that is defined to distinguish watermarked data from unwatermarked data.
Ideally, the design of $\tilde{\mathbf{X}}$ and $D$ should make $\tilde{\mathbf{X}}$ distinguishable from $\mathbf{X}$ or other non-watermarked datasets using the same random key $\mathbf{e}$. This property ensures that we can effectively detect the presence of the watermark in the watermarked data. 
Notice that, the detection of watermarks in discrete (or mixture) data is generally challenging, regardless of the insertion methods. This is because of the identifiability issue.
Specifically, for the $j$th variable of individual $i$, when both $X_{ij}$ and $\tilde{X}_{ij}$ take the same value, it becomes impossible to determine which one corresponds to a given random key. 
This identifiability issue is particularly severe when the dataset contains only a single discrete variable.
In fact, existing methods are usually not able to handle such a case. To improve the identifiability of watermarks, multivariate settings are typically required. In Section \ref{sec:method}, we show that the proposed method is capable of detecting a single discrete variable, due to the use of continualization.
Furthermore, the watermark detection procedures are expected to be robust against certain modifications. For example, the watermark should ideally remain detectable, even if 
one uses only
a subset of the watermarked data $\tilde{\mathbf{X}}$.

In addition to watermark detection, certain detection methods can also support \emph{user attribution} \citep{lu2025wasa,zhang2024personamark}. That is, the ability to attribute each watermark to its users. Let $A(Z_i,e_k)$ be the attribution function that determines whether $Z_i$ can be attributed to individual $k$. Then $A(Z_i,e_k)$ serves as a decision rule for identifying whether $Z_i$ originates from individual $k$:
\begin{equation*} 
    A(Z_i,e_k)=\begin{cases}
1 \quad (Z_i \text{ is attributed to user } k), & \text{if } (Z_i,e_k)\in C_k \\
0 \quad (Z_i \text{ is not attributed to user } k), & \text{otherwise,}
\end{cases}
\end{equation*}
where $C_k$ is the acceptance region that is uniquely associated with individual $k$, $k=1,\ldots,n$. In fact, most existing studies and real-world applications of user attribution focus on text and image watermarking. In Section \ref{sec:detection}, we show that the proposed method is also able to achieve user attribution for tabular data.

\subsection{Background: Distribution-Invariant Transformation} \label{sec:dip}
In this subsection, we briefly review the Distribution-Invariant Privatization (DIP) method \citep{bi2023distribution}, which motivates the design of the proposed watermark insertion method.
Let $X_1,X_2,\ldots,X_n$ be a sample from a known univariate continuous distribution $F$.
The DIP procedure proceeds as follows. For each $X_i$, $i=1,\ldots,n$, we have $F(X_i)$ follow $\mathrm{Uniform}(0,1)$. We then add a Laplace noise $e_i$ to $F(X_i)$ and obtain $F(X_i)+e_i$, where $e_i\overset{iid}{\sim} \mathrm{Laplace}(0,b)$ for a known value $b>0$. Let $G$ be the CDF of $\mathrm{Uniform}(0,1)$ plus $\mathrm{Laplace}(0,b)$, which can be explicitly derived through convolution. Then 
$G(F(X_i)+e_i)$ follows $\mathrm{Uniform}(0,1)$ again. Finally, we apply $F^{-1}$ and
\begin{equation} \label{DIP}
    \tilde{X}_i := F^{-1}\circ G(F(X_i)+e_i)
\end{equation}
follows the original distribution $F$. The selection of a Laplace distribution is to satisfy the differential privacy definition \citep{10.1007/11787006_1}. In the watermarking context, however, such a requirement is not necessary.

When data is discrete, 
DIP subtracts a uniformly distributed random variable $U_i$ from $X_i$, i.e. $X_i-U_i$, such that the resulting distribution
becomes continuous.
For example, when $X_i$ takes integer values, it can use $U_i \stackrel{iid}{\sim} \mathrm{Uniform}(0,1)$. 
Similar transformations are also provided for the continualization of 
mixture variables. 
Then after applying \eqref{DIP}, a ceiling function corresponding to the specific continualization method can be applied. This converts the perturbed data back to its original domain, which also follows its original data distribution. The same technique can also be applied to categorical variables, after encoding them as multiple dummy variables.
See \citep{bi2023distribution} for more details.
In the following sections, unless otherwise stated, the variables to be watermarked are assumed to be continuous.

Realistically, the true distribution $F$ is usually unknown, which makes the direct application of \eqref{DIP} challenging. Bi and Shen \citep{bi2023distribution} proposed some practical solutions. However, those solutions are
not designed for the watermarking purpose, and there is no corresponding detection method to identify the embedded noise in DIP. These motivate the design of the proposed method as illustrated in the following section.

\section{Method} \label{sec:method}
%

In this section, we present the proposed STAMP, a watermarking framework for tabular data that accommodates numerical, discrete (including ordinal and categorical), and mixed data types while (asymptotically) preserving their underlying distribution. In Section \ref{sec:insertion}, we introduce the proposed insertion method, which adapts the design of DIP as introduced in Section \ref{sec:dip} to the context of watermarking. In Section \ref{sec:detection}, we introduce the proposed detection method, which is completely novel.


\subsection{Insertion Method} \label{sec:insertion}


Suppose $\mathbf{X}_i=(X_{i1},X_{i2},\ldots,X_{ip})$ is a $p$-dimensional random vector, whose joint distribution is $F$. 
Let $\mathbf{e}_i = (e_{i1}, e_{i2}, \ldots, e_{ip})$ denote the \emph{key} to be inserted into $\mathbf{X}_i$, where $e_{ij} \stackrel{iid}{\sim} \mathrm{Laplace}(0,b)$ for all $j \in \{1,\ldots,p\}$. And let $\mathbf{W} = (W_1,\ldots,W_p)$ be the corresponding insertion functions. For $W_1$, we apply the univariate transformation as in \eqref{DIP} to $X_{i1}$. That is, 
\begin{equation} \label{eq:dim_1}
    \tilde{X}_{i1} = W_1(X_{i1},e_{i1}) := F_1^{-1}\circ G(F_1(X_{i1})+e_{i1}),
\end{equation}
where $F_1$ is the marginal distribution of $X_{i1}$. 
For notational simplicity, let $F_j = F_{X_j|X_1,\ldots,X_{j-1}}$ be the conditional distribution of $X_j$ given $(X_1,\ldots,X_{j-1})$. Then for $j=2,\ldots,p$, we follow the probability chain rule and apply
\begin{equation} \label{eq:dim_j}
\begin{aligned}
\tilde{X}_{ij}
&= W_j(X_{ij},e_{ij}) \\
&:= F^{-1}_{j}
\Big(
G\big(
F_{j}
(X_{ij} \mid X_{i1},\ldots,X_{i(j-1)})
+ e_{ij}
\big)
\;\Big|\;
\tilde{X}_{i1},\ldots,\tilde{X}_{i(j-1)}
\Big).
\end{aligned}
\end{equation}
Following Section \ref{sec:dip} and the probability chain rule, it can be seen that $\tilde{\mathbf{X}}_i=(\tilde{X}_{i1},\ldots,\tilde{X}_{ip})$ also follows $F$.


Here the \emph{key} $\mathbf{e}$ will be used in the subsequent detection method to identify if $\tilde{\mathbf{X}}$ has been watermarked. In fact, $\mathbf{e}$ can follow any known distribution as long as we change the function $G$ accordingly. We adopt the Laplace distribution because of its heavy-tail nature, which can potentially improve watermark detection rate. 

The proposed watermark insertion algorithm can be summarized as follows:
\begin{algorithm}[H]
\caption{STAMP Watermark Insertion Algorithm} \label{alg:insertion}
\footnotesize
\begin{algorithmic}[1]
\STATE \textbf{Input:} Original data $\{\mathbf{X}_i\}_{i=1}^n$, noise scale $b>0$
\STATE \textbf{Output:} Watermarked data $\{\tilde{\mathbf{X}}_i\}_{i=1}^n$

\FOR{$j=1$ \TO $p$}
  \FOR{$i=1$ \TO $n$}
    \STATE Draw $e_{ij}\sim \mathrm{Laplace}(0,b)$
    \STATE $\tilde{X}_{ij}=W_j(X_{ij},e_{ij})$, according to \eqref{eq:dim_1} for $j=1$ and \eqref{eq:dim_j} for $j>1$
  \ENDFOR
\ENDFOR
\RETURN $\{\tilde{\mathbf{X}}_i\}_{i=1}^n$
\end{algorithmic}
\end{algorithm}

For a discrete (and mixture) variable $j$, we first apply continualization as described in Section \ref{sec:dip}, followed by Algorithm \ref{alg:insertion}. Notably, once Algorithm \ref{alg:insertion} is completed, we skip the re-discretization step and return the raw, continuous output $\tilde{X}_{ij}$ to avoid the identifiability issue when both $X_{ij}$ and $\tilde{X}_{ij}$ take the same value. On the one hand, the end user who requests watermarking can easily apply a ceiling function and get a discrete variable for their own application. On the other hand, the raw output $\tilde{X}_{ij}$ is more informative than ceiling$(\tilde{X}_{ij})$ for recovering the watermark key $e_{ij}$, which helps distinguish $\tilde{X}_{ij}$ from $X_{ij}$ or other data. This design also allows us to watermark and detect even one-dimensional discrete variables.

In practice, the true distribution $F$ is usually unknown. 
One possible solution is to use kernel smoothing, which may be computationally intensive, especially in high-dimensional settings. Alternatively, one may estimate $F$ using the empirical distribution function (EDF). However, the EDF usually has a bounded support and is generally not invertible. Therefore, we propose a refined EDF, denoted by $\hat{F}$, to estimate $F$. It intends to accommodate the need
for both insertion and detection of watermarks. Following our insertion method in Equations \eqref{eq:dim_1} and \eqref{eq:dim_j}, we construct $\hat{F}$ by sequentially defining $\hat{F}_1,\ldots,\hat{F}_p$, corresponding to $F_1,\ldots,F_p$.

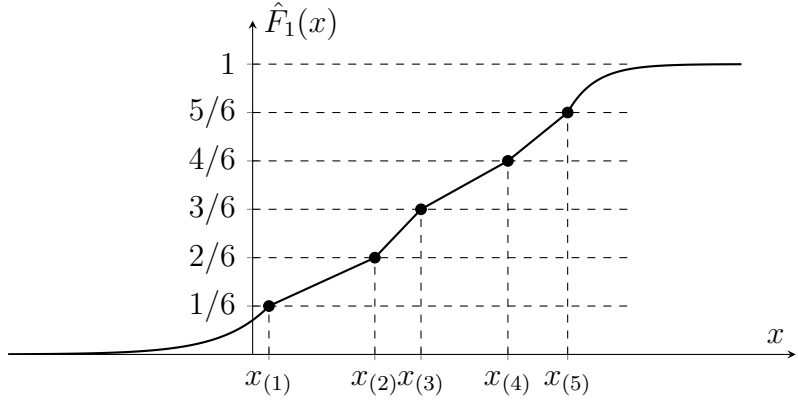
\begin{figure}
\centering
\begin{tikzpicture}
\begin{axis}[
    width=12cm,
    height=6cm,
    axis lines=middle,
    xmin=-4.5, xmax=10,
    ymin=0, ymax=1.15,
    xtick={0.3,2.25,3.1,4.7,5.8},
    xticklabels={$x_{(1)}$,$x_{(2)}$,$x_{(3)}$,$x_{(4)}$,$x_{(5)}$},
    ytick={1/6,2/6,3/6,4/6,5/6,1},
    yticklabels={$1/6$,$2/6$,$3/6$,$4/6$,$5/6$,$1$},
    xlabel={$x$},
    ylabel={$\hat F_1(x)$},
    ylabel style={yshift=0.4cm},
    samples=300,
]

\addplot[
    black,
    thick,
    domain=-4.5:0.3
]
{(1/6)*exp(1.15*(x-0.3))};

\addplot[black, thick, domain=0.3:2.25]
{1/6 + (1/6)*(x-0.3)/(2.25-0.3)};

\addplot[black, thick, domain=2.25:3.1]
{2/6 + (1/6)*(x-2.25)/(3.1-2.25)};

\addplot[black, thick, domain=3.1:4.7]
{3/6 + (1/6)*(x-3.1)/(4.7-3.1)};

\addplot[black, thick, domain=4.7:5.8]
{4/6 + (1/6)*(x-4.7)/(5.8-4.7)};

\addplot[
    black,
    thick,
    domain=5.8:9
]
{1 - (1/6)*exp(2*(5.8-x))};

 \addplot[dashed, domain=-0:7] {1/6};
 \addplot[dashed, domain=-0:7] {2/6};
 \addplot[dashed, domain=-0:7] {3/6};
 \addplot[dashed, domain=-0:7] {4/6};
 \addplot[dashed, domain=-0:7] {5/6};
 \addplot[dashed, domain=-0:7] {1};

 \addplot[dashed] coordinates {(0.3,0) (0.3,1/6)};
\addplot[dashed] coordinates {(2.25,0) (2.25,2/6)};
 \addplot[dashed] coordinates {(3.1,0) (3.1,3/6)};
 \addplot[dashed] coordinates {(4.7,0) (4.7,4/6)};
 \addplot[dashed] coordinates {(5.8,0) (5.8,5/6)};

\addplot[
    only marks,
    mark=*,
    mark size=2pt
]
coordinates {
    (0.3,1/6)
    (2.25,2/6)
    (3.1,3/6)
    (4.7,4/6)
    (5.8,5/6)
};

\end{axis}
\end{tikzpicture}
\caption{Illustration of the proposed refined empirical distribution function in the one-dimensional case, where $n=5$ distinct values are observed.}
\label{fig:f1}
\end{figure}

Specifically, for $F_1$, 
we introduce two tail extensions to the empirical distribution function, so that its domain covers the entire real line. We then interpolate between the probability masses of the EDF to obtain a continuous and strictly monotonic function. This ensures the invertibility of $\hat{F}_1$ on the whole real line. 
Let $x_{(1)}<\dots<x_{(n)}$ be the distinct ordered values observed in the sample $X_{11},\dots,X_{n1}$. Here the underlying distribution is assumed to be either continuous or obtained by continualizing a discrete distribution. Therefore, ties occur with probability zero and all observed values are distinct. Then we define the refined empirical distribution function as:
$$
\hat{F}_1(x)=\begin{cases}
\frac{1}{n+1}e^{x-x_{(1)}} & x < x_{(1)} \\
\frac{1}{n+1}(i+\frac{x-x_{(i)}}{x_{(i+1)}-x_{(i)}}) & x_{(i)} \leq x < x_{(i+1)} \\
1-\frac{1}{n+1}e^{x_{(n)}-x} & x \geq x_{(n)}.
\end{cases}
$$
Figure \ref{fig:f1} shows an example of the refined empirical distribution function.

For the conditional distributions $F_j$, $j = 2,\ldots,p$, 
we find the observation $\mathbf{x}^*$ in the dataset whose first $j-1$ coordinates are closest to the target point. We then construct the EDF using the $j$th coordinate of $\mathbf{x}^*$ (as $x_{(2)}$) together with the largest distinct observed value in the dataset that is smaller than this coordinate (as $x_{(1)}$). Similar to the definition of $\hat{F}_1$, we introduce two tails to the empirical CDF and interpolate between the probability masses. This ensures the invertibility of $\hat{F}_j$ on the whole real line. The formula is as follows:
$$
\hat{F}_j(x)=\begin{cases}
\frac{1}{q}e^{x-x_{(1)}} & x < x_{(1)} \\
\frac{1}{q}+\frac{(1-\frac{2}{q})(x-x_{(1)})}{x_{(2)}-x_{(1)}} & x_{(1)} \leq x < x_{(2)} \\
1-\frac{1}{q}e^{x_{(2)}-x} & x \geq x_{(2)}.
\end{cases}
$$
Here we set $\frac{1}{q}=O(\frac{1}{n})$. Theoretically, the resulting $\hat{F}$ remains consistent for the true distribution $F$. As a result, the output $\{\tilde{\mathbf{X}}_i\}_{i=1}^n$ of Algorithm \ref{alg:insertion} also follows $F$ asymptotically. Please see Section \ref{sec:theory} for more details.


There are some important considerations in the estimation of $\hat{F}$.
One is the choice of exponential tails. First, the exponential tails cover the entire real line, making detection relatively more feasible than tails with bounded support. Second, the exponential tails have a simple form, which makes its inverse function easy to derive and implement. Third, the exponential tails converge to zero at an exponential rate, allowing the refined EDF to remain close to the true distribution and converge to $F$ at a faster rate.
Nevertheless, alternative tail distributions may also be used and could lead to improved insertion and detection performance on certain datasets.
Another potential extension is to incorporate privacy-preserving mechanisms into the construction of $\hat{F}$. To protect the privacy of the data used in constructing $\hat{F}$, one may consider kernel smoothing to remove the non-differentiability induced by individual data points. Alternatively, one may use a differentially private EDF estimation (e.g., \citep{he2023algorithmically}), which perturbs the corresponding histogram to reduce the likelihood of data leakage.

\subsection{Detection Method} \label{sec:detection}
In this subsection, we describe the proposed detection procedure.
Intuitively, detection uses the estimated CDF $\hat{F}$ to identify the inserted key $\mathbf{e}$.

Following Section \ref{sec:insertion}, let $\mathbf{X} = (\mathbf{X}_1,\ldots,\mathbf{X}_n)'$ be the original dataset, and $\tilde{\mathbf{X}}$ its watermarked counterpart. Let $\mathbf{Z}$ be an $m \times p$ candidate dataset. Our goal is to determine whether $\mathbf{Z}$ is the watermarked original dataset or a subset thereof. Accordingly, we allow $m \le n$. That is, we aim to detect whether the watermark proposed in Section \ref{sec:insertion} is present in $\mathbf{Z}$. We assume that $\mathbf{Z}$ is either entirely watermarked or entirely unwatermarked. (Partial watermarking at the row level is associated with user attribution, which is discussed in Remark \ref{rem:attribution}.) 

We start with the simplest case. 
Let $\mathbf{Z} = (Z_1, Z_2, \dots, Z_m)'$ denote a sample of univariate continuous data. Our goal is to determine whether this sample has been watermarked using the insertion method in \eqref{eq:dim_1}. 
Specifically, we first apply $G^{-1}\circ \hat{F}$ to each $Z_i$, $i=1,\ldots,m$. If $Z_i$ is watermarked, i.e., $Z_i=\hat{F}^{-1}\circ G(\hat{F}(X_i)+e_i)$, then the transformation yields $\hat{F}(X_i)+e_i$. If $Z_i$ is not watermarked, however, the resulting values will generally not take this form. Note that $\hat{F}(X_i) \in [0,1]$ and $e_i$ is from our saved noise sequence $\mathbf{e}$. Hence, a necessary condition for $Z_i$ to be a watermarked version of $X_i$ is that $G^{-1}\circ \hat{F}(Z_i) - e_i \in [0,1]$. 

Practically, data may be shuffled before detection. And it is rather hard to know which $X_k$, $k \in \{1,2,\ldots,n\}$, can match with $Z_i$ after shuffling. In other words, if $Z_i$ is watermarked, then there exists $k \in \{1,2,\ldots,n\}$, such that $G^{-1}\circ \hat{F}(Z_i) - e_k \in [0,1]$. For unwatermarked data, this cannot be guaranteed. Motivated by this observation, we propose to use the following event as the watermark detection rule for each $Z_i$: 
\begin{equation} \label{eq:individual_detection_rule}
    \bigcup_{k=1}^n \Big\{G^{-1}\circ \hat{F}(Z_i) - e_k \in [0,1]\Big\}.
\end{equation}
Below we provide some intuition for \eqref{eq:individual_detection_rule}. On the one hand, we have
\begin{equation} \label{eq:necessary}
    Z_i \mbox{ is watermarked} \quad \Rightarrow \quad \bigcup_{k=1}^n \Big\{G^{-1}\circ \hat{F}(Z_i) - e_k \in [0,1]\Big\}.
\end{equation}
On the other hand, the converse implication also holds in probability, and the probability goes to 1 as $b \rightarrow \infty$. In other words, as the signal strength of the inserted key increases, the proposed watermark becomes easier to detect. Intuitively, for watermarked data, we always have $G^{-1}\circ \hat{F}(Z_i) - e_k \in [0,1]$ for at least one $k$, regardless of the value of $b$. 
Meanwhile, we note that for any given $Z_i$, 
the probability of $G^{-1}\circ \hat{F}(Z_i) - e_k \in [0,1]$ is bounded by $\frac{1}{2b}$ and goes to zero as $b$ goes to infinity.
That is, when $Z_i$ is not watermarked, 
we have 
\begin{equation} \label{eq:sufficient}
P\!\left(
\bigcup_{k=1}^n 
\Big\{G^{-1}\circ \hat{F}(Z_i) - e_k \in [0,1]\Big\}
\right)
\;\longrightarrow\; 0, \mbox{ as }
 b \to \infty.
\end{equation}
Here choosing $e_i$ to follow a Laplace distribution makes the probability in \eqref{eq:sufficient} smaller than under lighter-tailed distributions such as the Gaussian distribution.
By combining \eqref{eq:necessary} and \eqref{eq:sufficient}, we declare $Z_i$ to be watermarked if and only if \eqref{eq:individual_detection_rule} holds.
For discrete and mixture data, the detection procedure can be defined analogously in principle. Following the discussion in Section \ref{sec:insertion}, we require the continuous version of each $Z_i$ to avoid the identifiability issue. Subsequently, we also apply \eqref{eq:individual_detection_rule} to detect watermarks.

For the entire dataset $\mathbf{Z}$, we notice that if $\mathbf{Z}$ is watermarked, then every $Z_i$ can be matched to some $k \in \{1,\ldots,n\}$ such that \eqref{eq:individual_detection_rule} holds. And similar to \eqref{eq:sufficient}, the converse statement also holds in probability as $b\to\infty$. Therefore, we propose the following decision rule for univariate data, and declare that $\mathbf{Z}$ is watermarked if and only if:
$\bigcap_{i=1}^m\bigcup_{k=1}^n \Big\{G^{-1}\circ \hat{F}(Z_i) - e_k \in [0,1]\Big\}.$
The theoretical characterization and proof of our statements above can be found in Section \ref{sec:theory}. The result also holds when $n$ goes to infinity.

For multivariate data, the detection outcome depends on the desired decision rule. In our setting, we declare a dataset $\mathbf{Z}$ as watermarked if all variables are identified as watermarked. This criterion reflects a practical perspective: an unauthorized use of the whole original dataset should be regarded as infringement.\footnote{For alternative watermarking decision rules (e.g., declaring a dataset as watermarked if at least one variable is identified as watermarked), the subsequent rule \eqref{eq:multivariate_dataset_detection_rule} can be adjusted accordingly.} Mathematically, let $\mathbf{Z} = (\mathbf{Z}_1, \mathbf{Z}_2, \dots, \mathbf{Z}_m)'$ and $\mathbf{Z}_i = (Z_{i1},Z_{i2},\ldots,Z_{ip})$ for each $i=1,\ldots,m$. Then we declare that $\mathbf{Z}$ is watermarked if and only if:
\begin{equation} \label{eq:multivariate_dataset_detection_rule}
    \bigcap_{j=1}^p\bigcap_{i=1}^m\bigcup_{k=1}^n \Big\{G^{-1}\circ \hat{F}_j(Z_{ij}) - e_{kj} \in [0,1]\Big\}.
\end{equation}
In other words, we use the conditional distribution $\hat{F}_j$, which was used in the insertion step, to sequentially detect if the corresponding variable is watermarked.
Here we assume that the key for each variable is stored separately, such that each $Z_{ij}$ is matched only with $\{e_{kj}\}_{k=1}^n$ and not with $\{e_{kj'}\}_{k=1}^n$ for $j' \neq j$.


Then we can summarize a general detection algorithm as follows:
\begin{algorithm}[H]
\caption{STAMP Watermark Detection Algorithm} \label{alg:detection}
\footnotesize
\begin{algorithmic}[1]
\STATE \textbf{Input:} Given data $\{\mathbf{Z}_k\}_{k=1}^m$, saved key $\{\mathbf{e}_i\}_{i=1}^n$ and $\hat{F}$ 
\STATE \textbf{Output:} Detection result: 1 (watermarked) or 0 (not watermarked)
     \IF {\eqref{eq:multivariate_dataset_detection_rule} is satisfied}  
    \RETURN 1
    \ELSE \RETURN 0
    \ENDIF
\end{algorithmic}
\end{algorithm} 

Notice that, the proposed detection method is achieved for each $\mathbf{Z}_i$ individually. This allows it to be used even when we only have a very small $m$ (e.g., $m=1$). 
This is illustrated in our numerical studies in Sections \ref{sec:simulation} and \ref{sec:real_data}. In the meantime, this also allows parallel computing, which can improve the computational speed in a large-scale data setting.

\begin{rem}[User Attribution] \label{rem:attribution}
    \normalfont
    In addition to watermark detection, the proposed STAMP method also enables watermark-based user attribution. Notice that \eqref{eq:necessary} and \eqref{eq:sufficient} are directly applicable to each individual. 
    Then for each individual $i \in \{1,\ldots,m\}$, we can use $$\bigcup_{k=1}^n \bigcap_{j=1}^p\Big\{G^{-1}\circ \hat{F}_j(Z_{ij}) - e_{kj} \in [0,1]\Big\}$$ to match $\mathbf{Z}_i$ to a \emph{user-specific fingerprint} $\mathbf{e}_k$ in a coordinate-wise manner (i.e., requiring agreement across all variables). Consequently, this identifies a $k \in \{1,\ldots,n\}$ that uniquely links $\mathbf{Z}_i$ to individual $k$ in the original watermarked data $\tilde{\mathbf{X}}$.
    In the Appendix, we provide a more rigorous formulation of this attribution procedure, along with numerical studies demonstrating its effectiveness.
\end{rem}

One important consideration is the choice of $b$. A larger $b$ generally improves detection effectiveness while still preserving the marginal distribution. Meanwhile, a larger $b$ also weakens the relationship between the watermarked data and the original data, which may potentially reduce data utility. Therefore, choosing $b$ involves a practical trade-off.
On the one hand, $b$ should be large enough to ensure effective detection. On the other hand, among the values of $b$ that achieve satisfactory detection performance, one should prefer a relatively small value so that the watermarked data remains highly related to the original data. In our experiments, we select $b$ according to this principle by trying multiple candidate values and examining the corresponding performance. In fact, the selected $b$ can usually be reasonably small when achieving a good detection rate (e.g., $b=1$ or $2.5$). In practice, practitioners may follow the same principle when choosing $b$ for watermarked data release. When the key $\mathbf{e}$ follows a distribution other than Laplace, a scale parameter similar to $b$ can be selected in the same way, for example, the standard deviation in a Gaussian distribution.

\begin{rem}[No-Shuffle Detection] \label{rem:no_shuffle}
    \normalfont
    When the data is not shuffled (i.e., $X_i$ is paired with $\tilde{X}_i$), $\mathrm{Cor}(\tilde{\bm{X}},\bm{e})$ serves as an effective watermark detection method. Since $\tilde{X}_i=F^{-1}\circ G(F(X_i)+e_i)$, we expect a high correlation between $\tilde{\bm{X}}$ and $\bm{e}$, whereas the correlation between any unwatermarked $\bm{X}$ and $\bm{e}$ remains low due to their independence. 
\end{rem}

\section{Theoretical Properties} \label{sec:theory}
In this section, we study the theoretical properties of the proposed method. For the insertion procedure, we show that a distribution preservation property can be guaranteed, where the watermarked data follows the original distribution $F$ asymptotically. 
For the detection procedure, we focus on two key metrics: the true positive rate and the true negative rate, where a ``positive'' outcome indicates a watermarked dataset. Ideally, both should be close to one, so that we can reliably identify watermarked datasets and reduce susceptibility to watermark removal and forgery.
For our method, the true positive rate is always equal to one, as illustrated in \eqref{eq:necessary}. Consequently, our theoretical development primarily focuses on ensuring a high true negative rate under suitable asymptotic conditions. That is, when certain parameters (e.g. $n$, $b$) go to infinity, an unwatermarked dataset is correctly identified as unwatermarked with probability approaching one.

Specifically, for the insertion procedure, Theorem \ref{theorem3} shows that the watermarked data follows the original distribution $F$ 
when the sample size $n\to\infty$.

\begin{theorem}
The watermarked sample $\tilde{X}_1,...,\tilde{X}_n$ asymptotically follows $F$ as the size of the original sample $n \rightarrow \infty$.
    \label{theorem3}
\end{theorem}

For the detection procedure, we start from the univariate setting. For each $n$, let $X_1,\dots, X_n$ be i.i.d univariate continuous random variables which represent the original unwatermarked data.  
Define $f_n:\mathbb{R}\to \mathbb{R}$ as 
$f_n=G^{-1}\circ \hat{F},$ 
where $G$ and $\hat{F}$ follow the definition of Section \ref{sec:detection} and $\hat{F}$ is constructed from $X_1,\dots, X_n$. The subscript $n$ emphasizes the dependence of $f_n$ on the sample size through $\hat{F}$. 

\emph{Let $Z_1,\dots, Z_m$ denote a series of i.i.d univariate continuous samples which is not watermarked.} For discrete or mixed data, we follow Section \ref{sec:insertion} and assume that the raw continuous output of the insertion procedure is available. Let $F_{n1}$ denote the CDF of $f_n(Z_i)$, where $n$ indicates the dependence on the sample size and the subscript $1$ indicates that the distribution is associated with the first (and here, only) variable.
We then formally establish \eqref{eq:sufficient}, which justifies the watermark detection rule introduced in Section \ref{sec:detection}. 
Let $R_{ik}=\{f_n(Z_i)-e_k\notin[0,1]\}$, and $R=\bigcup_{i=1}^m \bigcap_{k=1}^n R_{ik}$. Then $R$ is the rejection region for watermark detection; that is, if $R$ occurs, we declare that the dataset $\mathbf{Z} = (Z_1,\dots, Z_m)'$ is not watermarked. Here, following the settings in Section \ref{sec:detection}, we have $m\le n$.
Lemma \ref{theorem1} provides conditions under which the false positive rate of our detection procedure (i.e., $\mathbf{Z}$ being misspecified as watermarked) can be controlled. Proofs of Lemma \ref{theorem1} and all subsequent theoretical properties are provided in the Appendix.

\begin{lemma}
     The following statements hold:
     \begin{enumerate}[(i)]
\item \textbf{Fixed \(n\).} Let \(e_1,\ldots,e_n \overset{\mathrm{i.i.d.}}{\sim} \mathrm{Laplace}(0,b)\). Then \(P(R)\to 1\) as \(b\to\infty\).

    \item \textbf{Diverging \(n\).} Let \(e_1,\ldots,e_n \overset{\mathrm{i.i.d.}}{\sim} \mathrm{Laplace}(0,b_n)\), where \(\{b_n\}_{n\ge 1}\) is a positive increasing sequence such that \(b_n\to\infty\). If \(n/b_n\to 0\), then \(P(R)\to 1\) as \(n\to\infty\).
     \end{enumerate}
     
  \label{theorem1}
\end{lemma}

Lemma \ref{theorem1} shows that, the detection rate converges to 1 either when $n$ is a constant or when  $n\to \infty$ and $b_n$ increases sufficiently quickly with $n$. Notice that, although Lemma \ref{theorem1} is stated under a Laplace distribution for $e_i$, similar results may also be derived if $e_i$'s follow a continuous distribution generally. Meanwhile, $Z_1,\dots, Z_m$ in Lemma \ref{theorem1} can represent any unwatermarked dataset. That is, the proposed method can asymptotically identify, with probability one, that a dataset has not undergone the proposed insertion procedure. 


Next, we further extend the conclusion of Lemma \ref{theorem1} by showing that the growth requirement on $b_n$ can be relaxed when $f_n(Z_i)$ satisfies certain mild conditions. Theorem \ref{theorem2} illustrates this result. 


\begin{theorem}
Following the setup in Lemma \ref{theorem1}, suppose that $F_{n1}$ is continuous with domain on the entire real line, there always exists a sequence $\{t_n\}$ such that $t_n \to +\infty$ as $n \to +\infty$, 
\(
nF_{n1}(-t_n)\to+\infty,
\)
and
\(
F_{n1}(-t_n)\to 0.
\)
If
\(
\frac{n}{b_n}\exp\left(-\frac{t_n}{b_n}\right)\to 0,
\)
then
\(
P(R)\to 1
\)
as $n\to+\infty$.
\label{theorem2}
\end{theorem}
  
Theorem \ref{theorem2} provides the possibility of a slower increase rate of $b_n$ under some restrictions on $F$. 
Compared with the result in Lemma \ref{theorem1}, the required growth rate of \(b_n\) is relaxed. Specifically, since \(t_n>0\), we have
$\frac{n}{b_n}\exp\left(-\frac{t_n}{b_n}\right)
<
\frac{n}{b_n}.$ and hence the condition
$\frac{n}{b_n}\exp\left(-\frac{t_n}{b_n}\right)\to 0$
is weaker than
$\frac{n}{b_n}\to 0.$

Notice that, when the data to be detected (e.g., $Z_1,\dots, Z_m$ in Lemma \ref{theorem1}) follow the same distribution as the original data, the detection task becomes particularly challenging. This happens when the data are either not generated by the proposed insertion procedure, or are generated using a different key. Corollary \ref{corollary1} shows that even under such a scenario, the proposed detection procedure can still correctly identify the data as unwatermarked. And in this case, no specific growth-rate requirement on \(b_n\) is needed.

  \newtheorem{corollary}{Corollary}
  
  \begin{corollary}
      Suppose $f_n(Z_i)\overset{d}{=} U+e$, where $U\sim \mathrm{Uniform}(0,1)$, $e\sim \mathrm{Laplace}(0,b_n)$ and $U$ and $e$ are independent. Then when $b_n\to +\infty$, $P(R)\to 1$ as $n\to +\infty$.
      \label{corollary1}
  \end{corollary}
  
Next, we extend our theoretical properties above to the multivariate case. Recall that in \eqref{eq:multivariate_dataset_detection_rule}, we declare a dataset as watermarked if all variables are identified as watermarked.
Let $\mathbf{Z}_i = (Z_{i1},Z_{i2},\ldots,Z_{ip})$, $i=1,\ldots,m$, denote a series of i.i.d. multivariate samples. 
Define $f_{n1},f_{n2},\ldots,f_{np}$ as a series of $\mathbf{R}\rightarrow \mathbf{R}$ functions, where $f_{nj}=G^{-1}\circ\hat{F}_{j}$, $j=1,\dots, p$. Here the definitions of $G$ and $\hat{F}_j$ follow Section \ref{sec:detection}. Then we denote $F_{nj}$ to be the cdf of $f_{nj}(Z_{ij})$. Let $R_{ijk}=\{f_{nj}(Z_{ij})-e_{kj}\notin[0,1]\}$ and $R_j=\bigcup_{i=1}^m \bigcap_{k=1}^n R_{ijk}$, as well as $R=\bigcup_{j=1}^p R_j$. Corollary \ref{corollary2} shows the control of the false positive rate for our detection procedure.
  \begin{corollary}
  The following statements hold:
     \begin{enumerate}[(i)]
    \item For any $F_{n1}, \dots, F_{np}$, 
      if $\frac{n}{b_n}\to 0$ as $n\to+\infty$, then $P(R)\to 1$.
     \item Suppose that, for some $j_0\in\{1,\dots,p\}$, $F_{nj_0}$ is continuous with domain on the entire real line, then there always exists a sequence $\{t_n\}$ such that $t_n \to +\infty$ as $n \to +\infty$,
\(
nF_{nj_0}(-t_n)\to+\infty,\
\)
and
\(
F_{nj_0}(-t_n)\to 0
\). If
\(
\frac{n}{b_n}\exp\left(-\frac{t_n}{b_n}\right)\to 0,
\)
then
$P(R)\to 1$ as $n\to\infty$.

  \end{enumerate}
    \label{corollary2}
  \end{corollary}



\section{Simulation Studies} \label{sec:simulation}

In this section, we conduct a series of simulation studies to evaluate the performance of the proposed watermarking framework under a variety of settings, including univariate and multivariate data, different data types, sample sizes, and subset levels.

We compare the proposed STAMP method with five existing competitive methods, including Green-List Watermark (GLW) \citep{he2024watermarking}, TabularMark \citep{zheng2024tabularmark}, TAB-DRW \citep{zhao2025tab}, TabWak \citep{zhu2025tabwak}, and MUSE \citep{fang2025muse}. GLW and TabularMark are adapted from text watermarking methods. They preserve statistical fidelity well, but are primarily designed for continuous data. TAB-DRW, TabWak, and MUSE rely on diffusion models for watermark insertion and detection. They can accommodate both continuous and discrete data, but require multivariate inputs and therefore cannot be applied to univariate data. Consequently, in the univariate setting, we compare STAMP only with GLW and TabularMark, whereas in multivariate settings when discrete variables are present, we compare it with TAB-DRW, TabWak and MUSE.

The evaluation of all competing methods includes three aspects: (i) watermark detection performance, measured by the detection rate; (ii) statistical fidelity of the generated data, quantified by the distributional difference between the watermarked and original data, and (iii) robustness to subsetting (i.e., detectability when only a subset of the original data is available). Due to the space constraints, most fidelity results are deferred to the Appendix.

\subsection{Univariate Case}


First, we conduct watermark insertion and detection for univariate data. We generate i.i.d samples from several commonly used distributions, including both continuous and discrete ones. The continuous distributions include $\mathrm{N}(0,1)$, $\mathrm{Exp}(1)$, $\mathrm{Uniform}(0,1)$, $\mathrm{Beta}(2,5)$, and $\mathrm{Laplace}(0,1)$, and the discrete distributions include $\mathrm{Poisson}(1)$, $\mathrm{Bin}(5,0.3)$, and $\mathrm{Bernoulli}(0.5)$. We vary the sample size $n \in \{500,2000,5000\}$. 

For the proposed STAMP method, we set the scale parameter $b$ of the inserted key as 1. 
For GLW, we set the split of interval $m=1000$, and significance level $\alpha=0.05$. For TabularMark, since only a subset of the samples is selected for watermarking, we set the number of selected samples $n_w=0.15n$, split of interval $k=500$, green-listed interval width $p=1$, and threshold for detection $1.96$. All parameter choices are based on the experiment settings used in the original papers. However, since neither of these two methods is able to watermark discrete data, they are excluded from comparisons in the discrete cases. 


For a fair comparison, all competing methods are applied to the same original dataset. That is, for each replicate, a watermark is embedded into the same dataset using each method, and the resulting watermarked data are then evaluated.
Each experiment is repeated 500 times.

For each method, the corresponding watermark insertion procedure is applied. The proposed method is compared with the competing methods based on three evaluation metrics, that is, detection, fidelity, and robustness against subsetting. In Sections \ref{sec:sim1_detection}, \ref{sec:sim1_robustness}, and \ref{sec:sim1_utility}, we present and discuss the results for each of these three aspects, respectively.

\subsubsection{Detection} \label{sec:sim1_detection}

For detection, we conduct the procedure introduced above for data generation and watermark insertion, then summarize the detection rate for evaluation. For each replicate, each method's detection procedure is applied to both its corresponding watermarked dataset and the original dataset.
We evaluate detection performance using the true positive (TP) rate and the true negative (TN) rate. The true positive rate is the proportion of watermarked datasets correctly identified as watermarked, and the true negative rate is the proportion of original datasets correctly identified as non-watermarked.

\begin{table}[!ht]
  \centering
  \footnotesize
  \setlength{\tabcolsep}{5.5pt}
  \renewcommand{\arraystretch}{1.15}
  \resizebox{0.7\textwidth}{!}{
  \begin{threeparttable}
  \caption{Detection results for continuous distributions across sample sizes.}
  \label{tab:table_detect_continuous}
  \begin{tabular}{@{}l l S[table-format=1.3] S[table-format=1.3]
                      S[table-format=1.3] S[table-format=1.3]
                      S[table-format=1.3] S[table-format=1.3]@{}}
    \toprule
    \multirow{2}{*}{Distribution} & \multirow{2}{*}{Method} &
      \multicolumn{2}{c}{$n=500$} &
      \multicolumn{2}{c}{$n=2000$} &
      \multicolumn{2}{c}{$n=5000$} \\
    \cmidrule(lr){3-4} \cmidrule(lr){5-6} \cmidrule(lr){7-8}
      & & {TP} & {TN} & {TP} & {TN} & {TP} & {TN} \\
    \midrule
    $\mathrm{N}(0,1)$     & STAMP         & 1.000 & 1.000 & 1.000 & 1.000 & 1.000 & 1.000 \\
                           & GLW         & 1.000 & 0.938 & 1.000 & 0.956 & 1.000 & 0.958 \\
                           & TabularMark & 1.000 & 0.984 & 1.000 & 0.980 & 1.000 & 0.980 \\
    \addlinespace[2pt]
    $\mathrm{Exp}(1)$      & STAMP         & 1.000 & 1.000 & 1.000 & 1.000 & 1.000 & 1.000 \\
                           & GLW         & 1.000 & 0.940 & 1.000 & 0.944 & 1.000 & 0.954 \\
                           & TabularMark & 1.000 & 0.984 & 1.000 & 0.980 & 1.000 & 0.980 \\
    \addlinespace[2pt]
    $\mathrm{Unif}(0,1)$   & STAMP         & 1.000 & 1.000 & 1.000 & 1.000 & 1.000 & 0.998 \\
                           & GLW         & 1.000 & 0.950 & 1.000 & 0.948 & 1.000 & 0.962 \\
                           & TabularMark & 1.000 & 0.984 & 1.000 & 0.980 & 1.000 & 0.980 \\
    \addlinespace[2pt]
    $\mathrm{Beta}(2,5)$   & STAMP         & 1.000 & 1.000 & 1.000 & 1.000 & 1.000 & 1.000 \\
                           & GLW         & 1.000 & 0.948 & 1.000 & 0.952 & 1.000 & 0.940 \\
                           & TabularMark & 1.000 & 0.984 & 1.000 & 0.980 & 1.000 & 0.980 \\
    \addlinespace[2pt]
    $\mathrm{Laplace}(0,1)$& STAMP         & 1.000 & 1.000 & 1.000 & 1.000 & 1.000 & 0.998 \\
                           & GLW         & 1.000 & 0.946 & 1.000 & 0.932 & 1.000 & 0.948 \\
                           & TabularMark & 1.000 & 0.984 & 1.000 & 0.980 & 1.000 & 0.980 \\
    \bottomrule
  \end{tabular}
  \begin{tablenotes}[flushleft]
    \footnotesize
    \item \textit{Notes:} Values are proportions computed from 500 replications. TP = true positive rate; TN = true negative rate.
  \end{tablenotes}
  \end{threeparttable}
  }
\end{table}

The results for the continuous distributions are reported in Table \ref{tab:table_detect_continuous}. STAMP achieves perfect true positive and true negative rates across all settings, regardless of the sample size or underlying distribution. The two competing methods also have high true positive rates, but their true negative rates are consistently lower than those of the proposed method. One possible reason is that the competing methods both adopt a hypothesis testing method for detection, which inherently has a default 5\% false positive rate. Since  detection is perfect for the proposed method across all sample sizes, no discernible trend in the true positive and true negative rates can be observed as the sample size increases.

Table \ref{tab:table_detect_discrete} reports the results for the discrete distributions. The proposed method achieves high true positive and true negative rates across different sample sizes and distribution types. As none of the competing methods is directly applicable to discrete data, no comparisons are made in these settings. Similar to the continuous case, no discernible trend in either metric can be observed as the sample size increases.

\begin{table}[!ht]
  \centering
  \footnotesize
  \setlength{\tabcolsep}{5.5pt}
  \renewcommand{\arraystretch}{1.15}
    \resizebox{0.55\textwidth}{!}{
  \begin{threeparttable}
  \caption{Detection results for discrete distributions across sample sizes.}
  \label{tab:table_detect_discrete}
  \begin{tabular}{@{}l S[table-format=1.3] S[table-format=1.3]
                      S[table-format=1.3] S[table-format=1.3]
                      S[table-format=1.3] S[table-format=1.3]@{}}
    \toprule
    \multirow{2}{*}{Distribution} &
      \multicolumn{2}{c}{$n=500$} &
      \multicolumn{2}{c}{$n=2000$} &
      \multicolumn{2}{c}{$n=5000$} \\
    \cmidrule(lr){2-3} \cmidrule(lr){4-5} \cmidrule(lr){6-7}
      & {TP} & {TN} & {TP} & {TN} & {TP} & {TN} \\
    \midrule
    $\mathrm{Poisson}(1)$    & 1.000 & 0.998 & 1.000 & 0.996 & 1.000 & 0.998 \\
    $\mathrm{Bin}(5,0.3)$ & 1.000 & 0.994 & 1.000 & 1.000 & 1.000 & 0.996 \\
    $\mathrm{Bernoulli}(0.5)$   & 1.000 & 0.974 & 1.000 & 0.964 & 1.000 & 0.982 \\
    \bottomrule
  \end{tabular}
  \begin{tablenotes}[flushleft]
    \footnotesize
    \item \textit{Notes:} Values are proportions computed from 500 replications. TP = true positive rate; TN = true negative rate.
  \end{tablenotes}
  \end{threeparttable}
  }
\end{table}

Meanwhile, we evaluate the changes that proposed watermark insertion has made to the original data. Table \ref{tab:distribution_results_transposed} in the Appendix shows the correlation between watermarked and unwatermarked data in various distributions. It can be seen that $\tilde{\bm{X}}$ remains highly similar to the original data across all settings.
In fact, the correlation values also depend on the choice of $b$. When $b$ is large, $\mathrm{Cor}(\bm{X},\tilde{\bm{X}})$ decreases due to the increased noise. In contrast, $\mathrm{Cor}(\tilde{\bm{X}},\bm{e})$ increases, which facilitates detection, as discussed in Remark \ref{rem:no_shuffle}. 
This reflects a trade-off between detection performance and similarity to the original data values.

\subsubsection{Robustness against Subsetting} \label{sec:sim1_robustness}
Next, we evaluate the robustness of the detection rate under subsetting. That is, we examine whether each watermark can still be detected when a user, either intentionally or unintentionally, uses only a subset of the watermarked dataset. We consider two subsetting scenarios: proportion-based and fixed-size sampling. In the first scenario, we randomly select 25\%, 50\% and 75\% of all observations. In the second scenario, we select only 1, 10, 100 observations, which are relatively small sizes in large-sample settings. For each case, the detection method is applied solely to the selected subset, and the detection rate is computed in the same manner as described earlier.

The results for the 25\%, 50\% and 75\% subsetting 
are reported in Tables \ref{tab:table_robustness_continuous_prop1}, \ref{tab:robustness_pi_050} and \ref{tab:robustness_pi_075} in the Appendix, respectively. Overall, STAMP maintains a high detection rate even as the proportion of selected observations decreases, demonstrating its robustness to subsetting. Among the competing methods, GLW also achieves consistently strong detection performance, whereas TabularMark performs poorly. This discrepancy may be attributed to its matching algorithm, which may be less effective when only a subset of the observations is available. 

The results for selecting 1 observation are reported in Table \ref{tab:robustness_prop_1}, while results for 10 and 100 observations are reported in Tables \ref{tab:robustness_prop_10} and \ref{tab:robustness_prop_100} in the Appendix, respectively. Among the competing methods, TabularMark continues to perform poorly in most cases, while GLW achieves satisfactory true positive rates when 10 or 100 observations are selected. However, when only one single observation is available, its performance deteriorates substantially. In contrast, STAMP maintains a high detection rate across all settings. In other words, \emph{in the extreme case where only one single observation is available, STAMP is still able to detect the underlying watermark.} This is attributed to the effective use of $\hat{F}$, as described in Algorithm \ref{alg:detection}.

We restrict the robustness evaluation to continuous distributions, as the competing methods are only applicable to continuous data. In an unreported study, we find that STAMP remains effective for detecting watermarks in discrete data (when the raw output of Algorithm \ref{alg:insertion} is used). A comparative evaluation would not be informative and is not included.

\begin{table}[!ht]
\centering
\footnotesize
\setlength{\tabcolsep}{5.5pt}
\renewcommand{\arraystretch}{1.15}
  \resizebox{0.7\textwidth}{!}{
\begin{threeparttable}
\caption{Robustness of watermark detection under subsetting for continuous distributions (1 observation selected).}
\label{tab:robustness_prop_1}
\begin{tabular}{@{}l l S[table-format=1.3] S[table-format=1.3]
S[table-format=1.3] S[table-format=1.3]
S[table-format=1.3] S[table-format=1.3]@{}}
\toprule
\multirow{2}{*}{Distribution} & \multirow{2}{*}{Method} &
\multicolumn{2}{c}{$n=500$} &
\multicolumn{2}{c}{$n=2000$} &
\multicolumn{2}{c}{$n=5000$} \\
\cmidrule(lr){3-4}\cmidrule(lr){5-6}\cmidrule(lr){7-8}
& & {TP} & {TN} & {TP} & {TN} & {TP} & {TN} \\
\midrule
$\mathrm{N}(0,1)$
& STAMP         & 1.000 & 1.000 & 1.000 & 1.000 & 1.000 & 1.000 \\
& GLW         & 0.000 & 1.000 & 0.000 & 1.000 & 0.000 & 1.000 \\
& TabularMark & 0.000 & 1.000 & 0.000 & 1.000 & 0.000 & 1.000 \\
\addlinespace[2pt]
$\mathrm{Exp}(1)$
& STAMP         & 1.000 & 1.000 & 1.000 & 1.000 & 1.000 & 1.000 \\
& GLW         & 0.000 & 1.000 & 0.000 & 1.000 & 0.000 & 1.000 \\
& TabularMark & 0.006 & 0.994 & 0.004 & 0.996 & 0.002 & 0.998 \\
\addlinespace[2pt]
$\mathrm{Unif}(0,1)$
& STAMP         & 1.000 & 1.000 & 1.000 & 1.000 & 1.000 & 1.000 \\
& GLW         & 0.000 & 1.000 & 0.000 & 1.000 & 0.000 & 1.000 \\
& TabularMark & 0.012 & 0.984 & 0.034 & 0.966 & 0.038 & 0.958 \\
\addlinespace[2pt]
$\mathrm{Beta}(2,5)$
& STAMP         & 1.000 & 1.000 & 1.000 & 1.000 & 1.000 & 1.000 \\
& GLW         & 0.000 & 1.000 & 0.000 & 1.000 & 0.000 & 1.000 \\
& TabularMark & 0.030 & 0.970 & 0.022 & 0.976 & 0.028 & 0.968 \\
\addlinespace[2pt]
$\mathrm{Laplace}(0,1)$
& STAMP         & 1.000 & 1.000 & 1.000 & 1.000 & 1.000 & 1.000 \\
& GLW         & 0.000 & 1.000 & 0.000 & 1.000 & 0.000 & 1.000 \\
& TabularMark & 0.004 & 0.996 & 0.000 & 1.000 & 0.000 & 1.000 \\
\bottomrule
\end{tabular}
\begin{tablenotes}[flushleft]
\footnotesize
 \item \textit{Notes:} Values are proportions computed from 500 replications. TP = true positive rate; TN = true negative rate.
\end{tablenotes}
\end{threeparttable}
}
\end{table}

\subsubsection{Fidelity} \label{sec:sim1_utility}

Lastly, we evaluate the fidelity of the dataset after watermark insertion. Specifically, we use the Kolmogorov–Smirnov (K–S) distance to measure the similarity between the original and watermarked data. For each replicate, we compute the K–S distance and then report its mean and standard deviation for each distribution, method, and sample size.

The results are reported in Table \ref{tab:table_ks_all} in the Appendix. We can see that STAMP maintains the K-S distance within a reasonable range for both continuous and discrete distributions. This indicates that the watermarked data remain close in distribution to the original data. Moreover, the K-S distance decreases as the sample size increases, suggesting the asymptotic consistency of the watermarked data. This observation is consistent with the theoretical results established in Theorem \ref{theorem3}.

Compared with other methods in the continuous setting, STAMP generally outperforms TabularMark in several cases but underperforms GLW in most cases. One possible reason is that the sample size of our method is still relatively small, meaning that the empirical distribution cannot approximate true distribution well, while GLW improves fidelity by increasing the split of interval $m$, which could be arbitrary and does not rely on the sample size. However, we note that GLW does not necessarily have asymptotic consistency. In other words,
there may exist a sufficiently large sample size beyond which the proposed method is expected to outperform GLW in terms of K-S distance. In the discrete setting, neither of the competing methods is applicable, whereas STAMP continues to show good performance.

\subsection{Multivariate Case: Logistic Regression} \label{sec:multi-variate-simu}
We also conduct simulation studies in a multivariate setting to demonstrate that STAMP can effectively handle mixed data consisting of both continuous and discrete variables. Specifically, we generate a binary response variable $Y$ and a $p$-dimensional covariate vector $\mathbf{X}$. The covariates consist of one continuous variable following a standard normal distribution and $p_0$ binary variables, where $p_0\in\{1,3,5\}$. Thus, the dimension of $\mathbf{X}$ is $p=p_0+1$, and the complete dataset contains $p_0+2$ variables including both $\mathbf{X}$ and $Y$.
The sample size is set to $n\in\{500,1000\}$. The response variable is generated according to the logistic regression model
$\mathrm{logit}(P(Y=1\mid \mathbf{X}))=\boldsymbol{\beta}^{\top}\mathbf{X},$
where $\boldsymbol{\beta}=(1,-1, 1, -1,\ldots,-1)^\top$ is a coefficient vector of length $p$. For all evaluations, the number of replications is set to $500$.

For STAMP, the Laplace scale parameter is set to $b=2.5$.
Since the simulated datasets contain discrete variables, GLW and TabularMark are not applicable. Therefore, we compare STAMP only with TabWak, MUSE, and TAB-DRW. Notice that, these competing methods require generative models to produce synthetic data. For each $(n,p)$ combination, we first generate a dataset according to the procedure described in the previous paragraph, and then train a diffusion model on it. To ensure a comparable computational budget across methods, all diffusion models are trained for 10 epochs.
After training, each generative model is used to generate 500 synthetic datasets, on which logistic regression models are fitted to evaluate utility and fidelity. For watermark detection, we follow the respective implementations in the original papers and use a significance level of $\alpha=0.05$ when applying their $Z$-score-based detection procedures.


Since the K--S distance is not suitable for multivariate distributions, we evaluate utility using the mean squared error between the estimated coefficient vector $\hat{\beta}$ obtained from the watermarked data and the true coefficient vector $\beta$, namely,
$\frac{1}{p}|\hat{\beta}-\beta|_2^2.$
Notably, STAMP preserves a one-to-one correspondence between each original observation $X_i$ and its watermarked counterpart $\tilde{X}_i$. This property enables observation-level fidelity evaluation and allows us to compute the Kullback-Leibler (K-L) divergence between the fitted probabilities based on the watermarked data and the true probabilities generated from the original model.
In contrast, TabWak, MUSE, and TAB-DRW generate entirely synthetic datasets. As a result, no observation-level correspondence exists between the original and generated data, making such fidelity measures, including the K-L divergence, inapplicable to these methods.

Moreover, we evaluate the robustness of STAMP against subsetting in the same multivariate setting. During the detection stage, we assume that only $1, 10, 100$ observations are available for watermark detection. In addition to the main setting where we set $b=2.5$, we also consider $b=32$, following the principles in Section \ref{sec:detection}, to validate our theoretical results (i.e., Corollary \ref{corollary2}) where the detection rate goes to 1 as $b$ increases. 
The rest of the watermark insertion procedure remains unchanged. 

Table \ref{tab:detection_selected_combined} shows the detection and robustness results for the settings $(n, p)=(500, 3)$ and $(1000, 5)$. 
The results for the remaining settings can be found in Table \ref{tab:md_detection_subset} in the Appendix.
STAMP maintains near-perfect true positive and true negative rates in all settings. Under subsetting, STAMP continues to perform well across all settings, whereas the competing methods either become inapplicable or exhibit substantially reduced true positive rates. We can also see that even under the main experimental setting with $b=2.5$, STAMP still achieves high true positive rates while maintaining reasonable true negative rates. It is also notable that as fewer observations are available, detection becomes more challenging, leading to performance degradation for all methods.

The results of fidelity are reported in Tables  \ref{tab:mse_results}, \ref{tab:kl_logistic_results} in the Appendix, from which we can see that STAMP consistently achieves the lowest MSE among all methods, as well as relatively low K-L divergence. This indicates that the watermarked data remain highly similar to the original data. Moreover, both metrics decrease as the sample size increases, which is consistent with the theoretical results established in Theorem \ref{theorem3}.

\begin{table}[htbp]
\centering
\footnotesize
\setlength{\tabcolsep}{4pt}
\renewcommand{\arraystretch}{1.15}

\resizebox{0.9\textwidth}{!}{
\begin{threeparttable}
\caption{Detection rates (TP and TN) for selected $(n,p)$ settings in the logistic regression setting.}
\label{tab:detection_selected_combined}
\begin{tabular}{ll cc cc cc cc cc}
\toprule
& & \multicolumn{2}{c}{TAB-DRW}
& \multicolumn{2}{c}{MUSE}
& \multicolumn{2}{c}{TabWak}
& \multicolumn{2}{c}{STAMP ($b=2.5$)}
& \multicolumn{2}{c}{STAMP ($b=32$)} \\
\cmidrule(lr){3-4}
\cmidrule(lr){5-6}
\cmidrule(lr){7-8}
\cmidrule(lr){9-10}
\cmidrule(lr){11-12}
$(n,p)$ & Subset Size & TP & TN & TP & TN & TP & TN & TP & TN & TP & TN \\
\midrule

\multirow{4}{*}{$(500,\,3)$}
& 1
& NA & NA
& 0.000 & 0.952
& 0.006 & 0.962
& 1.000 & 0.258
& 1.000 & 0.996 \\
& 10
& 0.012 & 0.532
& 0.000 & 0.952
& 0.006 & 0.962
& 1.000 & 0.310
& 1.000 & 1.000 \\
& 100
& 0.858 & 0.532
& 1.000 & 0.952
& 0.104 & 0.962
& 1.000 & 0.688
& 1.000 & 1.000 \\

& Full 
& 0.850 & 0.532
& 1.000 & 0.952
& 0.182 & 0.962
& 1.000 & 0.994
& 1.000 & 1.000 \\

\addlinespace

\multirow{4}{*}{$(1000,\,5)$}
& 1
& NA & NA
& 0.448 & 0.946
& 0.048 & 0.944
& 1.000 & 0.092
& 1.000 & 1.000 \\
& 10
& 0.198 & 0.896
& 0.424 & 0.946
& 0.054 & 0.944
& 1.000 & 0.110
& 1.000 & 1.000 \\
& 100
& 0.354 & 0.896
& 0.944 & 0.946
& 0.112 & 0.944
& 1.000 & 0.398
& 1.000 & 1.000 \\
& Full
& 0.002 & 0.896
& 1.000 & 0.946
& 0.446 & 0.944
& 1.000 & 0.992
& 1.000 & 1.000 \\

\bottomrule
\end{tabular}
\begin{tablenotes}
\footnotesize
\item Note: Values are detection rates computed from 500 replications. TP and TN denote the true positive rate and true negative rate, respectively. Subset size indicates the number of observations used for detection; ``Full'' corresponds to using the entire training set. STAMP ($b=2.5$) and STAMP ($b=32$) denote the settings where the scale parameter of the Laplace noise is set to 2.5 and 32, respectively.
\end{tablenotes}
\end{threeparttable}
}
\end{table}

\section{Real Data Application} \label{sec:real_data}

In this section, we apply STAMP to two real-world datasets, NHANES Diabetes Data \footnote{The NHANES Diabetes dataset is publicly available at \url{https://archive.ics.uci.edu/dataset/887}.} \citep{national_health_and_nutrition_health_survey_2013-2014_(nhanes)_age_prediction_subset_887} and Portuguese Bank Marketing Data \footnote{The Bank Marketing dataset is publicly available at \url{https://archive.ics.uci.edu/ml/datasets/bank+marketing}.} \citep{MORO201422}, to demonstrate its practical performance. 
The NHANES Diabetes dataset is collected from the National Health and Nutrition Examination Survey (NHANES) conducted by the National Center for Health Statistics (NCHS). The survey is designed to assess the health and nutritional status of adults and children in the United States through interviews and physical examinations during 2013-2014. The dataset contains information on various health indicators, including diabetes status, demographic characteristics, body measurements, blood pressure, and laboratory test results, comprising both continuous and discrete variables. We restrict our analysis to individuals with complete records, resulting in a sample of 2,277 individuals and evaluate the performance of the proposed method in a classification setting.

Similar to the simulation studies in Section \ref{sec:simulation}, we apply our insertion method to watermark this dataset and evaluate its performance in terms of fidelity, detection and robustness against subsetting. The proposed STAMP method is compared with Tab-DRW, TabWak, and MUSE, due to the presence of both discrete and continuous data. 

For the classification task, we randomly split the dataset into training and test sets in each replication, with 80\% of the data used for training and the remaining 20\% used for testing.
For all methods, we watermark the training set and train an XGBoost model to predict diabetes status using all other variables on the watermarked training set. Prediction performance is evaluated on the original test set. For STAMP, the scale parameter of the Laplace noise is set to $b=1$. 
For the competing methods, since the data split varies across replications, we retrain the generative model for each split to ensure that the test set is not used during model training. Specifically, for each replication, a new generative model is trained on the corresponding training set, and a synthetic watermarked dataset of the same size is generated. This synthetic dataset is then used to train an XGBoost model.
To remain consistent with the settings in Section \ref{sec:multi-variate-simu}, and to ensure comparable computational costs across methods, we train each generative model in the competing methods for only 10 epochs. 

For fidelity, we evaluate performance by comparing the AUC, accuracy, precision, recall, and F1 score of models trained on the original or watermarked training set. The evaluation of detection rates follows the same procedure as in Section \ref{sec:multi-variate-simu}.
To evaluate robustness to subsetting, we consider subset sizes of 1, 10, and 100. To ensure reliable watermark detection under such an extreme setting, we also consider a larger noise scale $b=16$, which is selected following the principles in Section \ref{sec:detection}, in addition to the baseline $b=1$. 

The detection results are reported in Table \ref{tab:nhanes_detection_subset_rate}. The advantage of STAMP is still evident. It achieves the highest detection rate when the full dataset is used. Under subsetting, STAMP ($b=1$) provides the best TP among all competing methods. Furthermore, when the noise scale is increased to $b=16$, the true negative rate is also improved to nearly 1 across all settings, including the extreme case when only one single observation is available for watermark detection. In addition, the fidelity results are reported in Table \ref{tab:nhanes_diabetes_xgb} in the Appendix. STAMP preserves the performance of the XGBoost model across all measures at a level comparable to that obtained using the original training data. This indicates that data watermarked by STAMP closely resemble the original data and maintain the fidelity for downstream data analysis.

In addition, we further evaluate our method on the Portuguese Bank Marketing dataset. The data description, experimental setup, and detailed results are provided in the Appendix.

\begin{table}[htbp]
\centering
\footnotesize
\setlength{\tabcolsep}{7pt}
\renewcommand{\arraystretch}{1.15}
\resizebox{0.9\textwidth}{!}{
\begin{threeparttable}
\caption{Watermark detection performance on the NHANES Diabetes dataset under different subset sizes.}
\label{tab:nhanes_detection_subset_rate}

\begin{tabular}{l *{5}{cc}}

\toprule
& \multicolumn{2}{c}{TAB-DRW} 
& \multicolumn{2}{c}{MUSE} 
& \multicolumn{2}{c}{TabWak} 
& \multicolumn{2}{c}{STAMP ($b=1$)} 
& \multicolumn{2}{c}{STAMP ($b=16$)} \\
\cmidrule(lr){2-3}\cmidrule(lr){4-5}\cmidrule(lr){6-7}\cmidrule(lr){8-9}\cmidrule(lr){10-11}

Subset Size & TP & TN & TP & TN & TP & TN & TP & TN & TP & TN \\
\midrule

1
& NA & 1.000
& 0.000 & 1.000
& 0.056 & 1.000
& 1.000 & 0.122
& 1.000 & 0.992

\\

10
& 0.058 & 1.000
& 0.474 & 1.000
& 0.030 & 1.000
& 1.000 & 0.764
& 1.000 & 1.000

\\

100
& 0.280 & 1.000
& 0.988 & 1.000
& 0.090 & 1.000
& 1.000 & 1.000
& 1.000 & 1.000

\\

Full
& 1.000 & 1.000
& 1.000 & 1.000
& 0.288 & 1.000
& 1.000 & 1.000
& 1.000 & 1.000

\\

\bottomrule
\end{tabular}

\begin{tablenotes}[flushleft]
\footnotesize
\item \textit{Notes:} Values are detection rates computed from 500 replications. TP and TN denote the true positive rate and true negative rate, respectively. Subset size indicates the number of observations used for detection; ``Full'' corresponds to using the entire training set. STAMP ($b=1$) and STAMP ($b=16$) denote the settings where the scale parameter of the Laplace noise is set to 1 and 16, respectively.
\end{tablenotes}
\end{threeparttable}
}
\end{table}

\section{Discussion} \label{sec:discussion}
In this paper, we propose a new data watermarking method. Watermark insertion is achieved through injecting noise (i.e., \emph{keys}) into the original data; and the inserted watermarks can be detected by tracing these keys and assessing whether they fall within a pre-defined detection region. 

The proposed method has four key advantages. 
First, the proposed method accommodates both univariate and multivariate data, as well as both continuous and discrete variables. This distinguishes it from most competing methods, which typically require either continuous data or multivariate structures. 
Second, both watermark insertion and detection in the proposed method are conducted at the individual observation level. This enables watermark detection even when the dataset under examination has as few as a single observation. In other words, in the extreme case, our method can determine whether \emph{a single value} (i.e., $n=1$ and $p=1$) is a watermarked version of an original data point. To the best of our knowledge, this capability is not demonstrated by existing methods.
Third, for the same reason, the proposed method naturally extends to user attribution. That is, in addition to detecting the presence of a watermark, it can also identify the specific individual to whom the watermarked data correspond.
Last, the proposed method is able to preserve the original data distribution, at least asymptotically. This property ensures that the watermarked data can be used with high fidelity. 


Theoretically, we show that the proposed method can achieve perfect detection and that the watermarked data follow the original distribution, both asymptotically. Meanwhile, through extensive simulation studies and real data applications, we demonstrate its effectiveness in terms of fidelity preservation, detection accuracy, and robustness to subsetting. Our method outperforms existing approaches in various scenarios. In addition, we empirically validate our method's capability for user attribution.

For future work, we plan to explore several directions. First, we aim to extend the proposed method to multi-modal data, which encompasses not only tabular data but also image, text and other forms of unstructured data. Second, we will investigate ways to improve the robustness of our watermarking method against numerous adversarial attacks, including data poisoning and adversarial sample insertion. 

\newpage
\section*{Supplementary Materials: Observation-Level Watermarking and Detection for Tabular Data}
\setcounter{section}{0}

\setcounter{figure}{0}

\setcounter{table}{0}
\renewcommand{\thefigure}{A\arabic{figure}}

\renewcommand{\thetable}{A\arabic{table}}

\renewcommand{\theequation}{A\arabic{equation}}

\renewcommand{\thesection}{A\arabic{section}}

\renewcommand{\thealgorithm}{A\arabic{algorithm}}
\section*{Appendix}

\section{Bank Marketing Data Analysis} \label{sec:bank_marketing}

We apply our method to a Bank Marketing Dataset to show its practical performance. This dataset consists of marketing campaign data collected from a Portuguese retail bank between 2008 and 2013 \citep{MORO201422}. This campaign aims to promote long-term deposits to potential clients through phone calls. During these calls, client's personal information is recorded, including age, education level, employment status, marital status, housing or personal loan status, and default status. Additionally, information such as device type, prior contact history related to the campaign, and whether the client is interested in subscribing to a term deposit (yes/no) is collected. The dataset contains 30,488 respondents with complete records. 

Because this dataset contains both continuous and discrete variables, only TAB-DRW, TabWak, MUSE, and the proposed method are applicable. 
To enable repeated experimentation and to reduce computational cost, we randomly select a subset of $n_s=2000$ observations from the full dataset in each replication, rather than using the full dataset only once. The number of replications is set to be 500. In each replication, each method is applied to the selected subset and undergoes the complete evaluation procedure, including watermark insertion, detection, and fidelity evaluation. 

Since the true $\beta$ is unknown, we measure fidelity using the mean squared error (MSE) and the mean absolute percentage error (MAPE) between $\hat{\beta}$ estimated from the watermarked data and $\hat{\beta}_o$ estimated from the full original Bank Marketing dataset. 
In addition, because the true data-generating probabilities are unknown, the K-L divergence between the fitted probability of the watermarked and the original dataset for our method is provided instead. 
For the competing methods, training 500 generative models is computationally prohibitive. Therefore, we instead train a single generative model for each competing method on the full dataset, running 1000 epochs for the generative models. The detection threshold $\alpha$ is also chosen to be 0.05. Notice that, this setup brings an additional advantage to the competing methods, as they are trained on the full dataset, whereas the proposed method operates only on the selected subset in each replication.
To assess robustness against subsetting, we further consider subsetting rates of 25\%, 50\% and 75\% when evaluating detection performance. The scale parameter of the Laplace noise is set to $b=1$. 

The results are reported in Table \ref{tab:bank_mse} and \ref{tab:bank_detection_subset_rate}.
From these tables, we can see that in terms of both MSE and MAPE, STAMP outperforms TAB-DRW but underperforms TabWak and MUSE. 
One possible reason is that, compared with the competing methods, our method only uses the selected subset instead of the full dataset. The selected subset may fail to capture the underlying data distribution. In contrast, the competing methods utilize the full dataset to train the diffusion model and generate synthetic data. In addition, STAMP still achieves relatively low K-L divergence.
For detection, the proposed method still has the highest detection rate across all settings, and the results also demonstrate strong robustness against $25\%, 50\%, 75\%$ subsetting. In contrast, TAB-DRW and TabWak have degraded true positive rates when the subset ratio is small. MUSE remains robust against subsetting; however, it maintains a true negative rate of approximately 95\% due to the use of hypothesis testing.

\begin{table}[htbp]
\centering
\small
\caption{Mean MSE (standard deviation), MAPE and K-L divergence using the Bank Marketing dataset.}
\label{tab:bank_mse}
\begin{tabular}{lccc}
\toprule
Method & MSE & MAPE & K-L Divergence\\
\midrule
TAB-DRW  & 16.63 (10.50) & 12.82 (9.58) & NA\\
MUSE     & 4.39 (7.47) & 5.67 (5.71) & NA\\
TabWak   & 3.48 (4,27) & 6.17 (1.50) & NA\\
STAMP & 9.15 (3.51) & 11.97 (21.87) & 0.049 (0.011)\\
\bottomrule
\end{tabular}
\end{table}

\begin{table}[htbp]
\centering
\small
\setlength{\tabcolsep}{5.5pt}
\renewcommand{\arraystretch}{1.15}
\begin{threeparttable}
\caption{Watermark detection performance on the Bank Marketing dataset under different subsetting ratios.}
\label{tab:bank_detection_subset_rate}

\begin{tabular}{l *{4}{S[table-format=1.3] S[table-format=1.3]}}
\toprule
& \multicolumn{2}{c}{TAB-DRW} & \multicolumn{2}{c}{MUSE} & \multicolumn{2}{c}{TabWak} & \multicolumn{2}{c}{STAMP} \\
\cmidrule(lr){2-3}\cmidrule(lr){4-5}\cmidrule(lr){6-7}\cmidrule(lr){8-9}
Subset ratio & {TP} & {TN} & {TP} & {TN} & {TP} & {TN} & {TP} & {TN} \\
\midrule
25\%  & 0.400 & 0.958 & 1.000 & 0.950 & 0.590 & 0.946 & 1.000 & 1.000 \\
50\%  & 0.576 & 0.958 & 1.000 & 0.950 & 0.684 & 0.946 & 1.000 & 1.000 \\
75\%  & 0.724 & 0.958 & 1.000 & 0.950 & 0.666 & 0.946 & 1.000 & 1.000 \\
100\% & 0.802 & 0.950 & 1.000 & 0.958 & 0.640 & 0.946 & 1.000 & 1.000 \\
\bottomrule
\end{tabular}

\begin{tablenotes}[flushleft]
\footnotesize
\item \textit{Notes:} Values are detection rates averaged over 500 replications. TP = true positive rate; TN = true negative rate.
\end{tablenotes}

\end{threeparttable}
\end{table}

\section{User Attribution}

As discussed in Remark \ref{rem:attribution}, the proposed detection method can identify the specific individual to whom the watermarked data are attributed. That is, 
for each individual, there is a unique fingerprint, through which the individual can be identified. 

Suppose the underlying dataset $\{\mathbf{X}_k\}_{k=1}^n$ consists of only continuous variable. Then for each $\mathbf{X}_k$,
the corresponding $\mathbf{e}_k=(e_{k1},\dots,e_{kp})$ is unique and can be used as its fingerprint. In contrast, for discrete variables, $\mathbf{e}_k$ alone may not be sufficient to identify the specific individual. This is because multiple individuals may have the same discrete value. For example, in the univariate binary case, we may have $k$ and $k'$ that both satisfy $X_k =1$ and $X_{k'} =1$. To address this issue, we also consider $\mathbf{U}_k = (U_{k1},\ldots,U_{kp})$, in addition to $\mathbf{e}_k$, to uniquely identify each individual. For $j=1,\ldots,p$, when the $j$th variable $X_{ij}$ is discrete (or mixed), $U_{kj}$ is the independent uniformly distributed variable  used for continualization as described in Section \ref{sec:dip}. Otherwise when $X_{kj}$ is continuous, we assign $U_{kj}=0$ (since no continualization is used). Therefore, in the general mixed-data setting, we can see that the pair $(\mathbf{U}_k,\mathbf{e}_k)$ can be uniquely associated with individual $k$, $k=1,\ldots,n$, and can thus be used as the individual $k$'s fingerprint.

To determine whether a given candidate row corresponds to the $k$-th individual in the dataset, we apply the $k$th individual's fingerprint $(\mathbf{U}_k,\mathbf{e}_k)$ to $\mathbf{x}$. 
Specifically, we examine whether 
$$G^{-1}\circ \hat{F}_{j}(x_j-U_{kj})-e_{kj} \in [0,1], \quad \mbox{for all }j=1,\ldots,p.$$
If this condition holds, then $\mathbf{x}$ is attributed to the $k$th individual, i.e., identified as the watermarked version of that individual's data in the original dataset. Equivalently, we declare that $\mathbf{x}$ is attributed to the $k$-th individual if
\begin{equation} \label{eq:user_attribution}
    \bigcap_{j=1}^p\ \Big\{G^{-1}\circ \hat{F}_j(x_j-U_{ki}) - e_{kj} \in [0,1]\Big\}.
\end{equation}
The complete user-attribution algorithm is summarized in Algorithm \ref{alg:attribution}.

\begin{algorithm}[htbp]
\caption{STAMP User Attribution Algorithm}
\footnotesize
\label{alg:attribution}
\begin{algorithmic}[1]
\STATE \textbf{Input:} Given data $\{\mathbf{X}_k\}_{k=1}^m$, saved fingerprints $\{(\mathbf{U}_i,\mathbf{e}_i)\}_{i=1}^n$ and $\widehat F$
\STATE \textbf{Output:} Attribution result: row index $j$ or NotWatermarked
\FOR{$k=1$ \TO $m$}
    \IF{\eqref{eq:user_attribution} is satisfied for $\mathbf{X}_k$ and some $(\mathbf{U}_i,\mathbf{e}_i)$}
        \STATE \textbf{return} $i$
    \ELSE
        \STATE \textbf{return} N/A (No user can be attributed)
    \ENDIF
\ENDFOR
\end{algorithmic}
\end{algorithm}

In theory, similar to the proofs of Lemma \ref{theorem1} and Corollary \ref{corollary2}, we can show that when the value of $b$ is large enough, each row can be attributed to the correct individual. Theorem \ref{theorem3} ensures that the distribution is still preserved after this transformation. 

Numerically, we use the Bank Marketing data as described in Section \ref{sec:bank_marketing} to evaluate the effectiveness of the proposed method. 
The experimental setup is identical to that in Section \ref{sec:bank_marketing}, except that the number of subset samples $n_s$ is set to be 200. The evaluation consists of two components. First, we assess the detection performance by reporting the true positive rate, i.e., the proportion of the watermarked dataset correctly identified as watermarked. This is also one of the evaluation metrics we considered in the main text.
Second, we evaluate user attribution by computing the proportion of the watermarked dataset that are correctly matched to their embedded fingerprints. The results are listed in Table \ref{tab:dip_detection_b}. We can see that our method can still maintain a perfect detection rate. In the meantime, it achieves high and stable user-attribution rate, even when the value of the scale parameter $b$ is as small as 0.5. To the best of our knowledge, none of the competing methods is designed for user attribution, and therefore no comparison is made.

\begin{table}[htbp]
\centering
\small
\setlength{\tabcolsep}{7pt}
\renewcommand{\arraystretch}{1.15}
\begin{threeparttable}
\caption{Detection and user-attribution rates using the Bank Marketing data.}
\label{tab:dip_detection_b}

\begin{tabular}{l *{4}{S[table-format=1.3]}}
\toprule
$b$ & {0.5} & {1} & {11} \\
\midrule
Detection & 1.000 & 1.000 & 1.000 \\
User-Attribution     & 0.700 & 0.694 & 0.697 \\
\bottomrule
\end{tabular}

\end{threeparttable}
\end{table}

\section{Proofs}
\subsection*{Proof of Theorem \ref{theorem3}}

We divide the proof into two parts corresponding to the univariate and multivariate settings.

\noindent\textbf{Part 1: Univariate Case}

Let $\tilde{F}$ denote the distribution of the watermarked sample $(\tilde{X}_1,\ldots,\tilde{X}_n)$ and let $F$ denote the distribution of the original sample. Let $\hat{F}_e$ be the empirical distribution function of the original sample. We construct the estimator $\hat{F}$ as
$$
\hat{F}(x)=
\begin{cases}
\frac{1}{n+1}e^{x-x_{(1)}} & x < x_{(1)}, \\[6pt]
\frac{1}{n+1}\left(i+\frac{x-x_{(i)}}{x_{(i+1)}-x_{(i)}}\right) & x_{(i)} \leq x < x_{(i+1)}, \\[8pt]
1-\frac{1}{n+1}e^{x_{(n)}-x} & x \geq x_{(n)} .
\end{cases}
$$

We first show that $\hat{F}$ is a consistent estimator of $F$ as $n\to\infty$, which in turn implies that $\tilde{F}$ is also consistent for $F$.

By the uniform convergence of the empirical distribution function under the Kolmogorov--Smirnov distance,
\[
\rho(\hat{F}_e,F)
=
\sup_z \left| \hat{F}_e(x)-F(x) \right|
=
O\!\left(\frac{1}{\sqrt{n}}\right)
\]
almost surely.

When $F$ is continuous, the construction of $\hat{F}$ implies that
\[
\rho(\hat{F},\hat{F}_e) \le \frac{2}{n+1}
\]
almost surely. Therefore, by the triangle inequality,
\[
\rho(\hat{F},F)
\le
\rho(\hat{F},\hat{F}_e)
+
\rho(\hat{F}_e,F)
=
O\!\left(\frac{1}{\sqrt{n}}\right)
\]
as $n\to\infty$.

When $F$ is discrete,
\[
\rho(\hat{F}_{V^*},F)
=
O\!\left(\frac{1}{\sqrt{n}}\right)
\]
by Lemma S1 of \citep{bi2023distribution}, where $\hat{F}_{V^*}$ denotes the empirical distribution function defined on $V^*$, i.e., the continualized version of $X$. As in the continuous case,
\[
\rho(\hat{F},\hat{F}_{V^*}) \le \frac{2}{n+1},
\]
which yields
\[
\rho(\hat{F},F)
\le
\rho(\hat{F},\hat{F}_{V^*})
+
\rho(\hat{F}_{V^*},F)
=
O\!\left(\frac{1}{\sqrt{n}}\right).
\]
Hence $\hat{F}$ is a consistent estimator of $F$.

Finally, recall that
\[
\tilde{X}_i = H(\hat{F}(X_i)+e_i),
\]
where
\[
H = L \circ \hat{F}^{-1} \circ G.
\]
Here $L$ is the identity mapping when $F$ is continuous, and $L=\bar{L}$ (the ceiling function) when $F$ is discrete. By the consistency of $\hat{F}$, the continuous mapping theorem, and the construction of $L$, we conclude that $\tilde{X}_i$ converges in distribution to $F$ as $n\to\infty$. Consequently,
\[
\tilde{F}(x) \to F(x)
\]
for each $z\in\mathbb{R}$.

Moreover, by the Glivenko--Cantelli theorem,
\[
\rho(F,\tilde{F}) \to 0
\]
almost surely as $n\to\infty$.

\noindent\textbf{Part 2: Multivariate Case}

We now establish the consistency of $\tilde{F}$ for $F$ in the multivariate setting. For simplicity, we assume that all variables are continuous.

Let $d_{jl}$ denote the $j$-th order statistic of the $l$-th coordinate of $(\tilde{X}_1,\ldots,\tilde{X}_n)$, where $j=1,\ldots,n$ and $l=1,\ldots,p$. Define
\[
d_{0l} = \min_{j=1,\ldots,n} d_{jl}-1 .
\]

Following the result of \citep{bi2023distribution}, suppose the marginal estimator is defined by
\[
\hat{F}_0(x_1)
=
\frac{1}{n}\frac{x_1-d_{k-1}}{d_k-d_{k-1}}
+
\frac{k-1}{n},
\qquad
x_1 \in (d_{k-1},d_k],\; k=1,\ldots,n,
\]
and the conditional estimator is defined as
\[
\hat{F}_0(x_l \mid x_1,\ldots,x_{l-1}) =
\begin{cases}
1, & x_l > d_{jl}, \\[6pt]
\dfrac{x_l-d_{j-1,l}}{d_{jl}-d_{j-1,l}}, 
& x_l\in(d_{j-1,l},d_{jl}], \quad l=2,\ldots,p, \\[10pt]
0, & x_l \le d_{j-1,l}.
\end{cases}
\]

The resulting joint estimator
$$
\hat{F}_0(x_1,\ldots,x_p)
=
\hat{F}_0(x_1)
\prod_{l=2}^p
\hat{F}_0(x_l \mid x_1,\ldots,x_{l-1})
$$
satisfies
\[
\rho(\hat{F}_0,F) \to 0
\]
as $n\to\infty$.

Next we compare $\hat{F}$ and $\hat{F}_0$. From their constructions, we have
\[
|\hat{F}(x_1)-\hat{F}_0(x_1)|
\le
\frac{1}{n+1},
\]
and for $l=2,\ldots,p$,
\[
|\hat{F}(x_l \mid x_1,\ldots,x_{l-1})
-
\hat{F}_0(x_l \mid x_1,\ldots,x_{l-1})|
\le
\frac{1}{q}.
\]

Therefore, since $\frac{1}{q}=O(\frac{1}{n})$
$$
\begin{aligned}
\hat{F}_0(x_1,\ldots,x_p)
&=
\hat{F}_0(x_1)
\prod_{l=2}^p
\hat{F}_0(x_l \mid x_1,\ldots,x_{l-1}) \\
&\le
\left(\hat{F}(x_1)+\frac{1}{n+1}\right)
\prod_{l=2}^p
\left(\hat{F}(x_l \mid x_1,\ldots,x_{l-1})+\frac{1}{q}\right) \\
&=
\hat{F}(x_1,\ldots,x_p)+O\!\left(\frac{1}{n}\right).
\end{aligned}
$$

Conversely, by the same argument,
\[
\hat{F}(x_1,\ldots,x_p)
\le
\hat{F}_0(x_1,\ldots,x_p)
+
O\!\left(\frac{1}{n}\right).
\]

Hence,
\[
\rho(\hat{F},\hat{F}_0)\to 0
\quad\text{as } n\to\infty.
\]

Combining this with the result of Part 1 yields
\[
\tilde{F}(x_1,\ldots,x_p) \to F(x_1,\ldots,x_p)
\quad\text{almost surely}.
\]
Following the argument of \citep{bi2023distribution}, it further follows that
\[
\rho(\tilde{F},F) \to 0 .
\]

This completes the proof.

\subsection*{Proof of Lemma \ref{theorem1}}

    For any fixed $i$ and $k$, note that
\[
R_{ik}^c
=
\{f_n(Z_i)-e_k\in[0,1]\}
=
\{e_k\in[f_n(Z_i)-1,f_n(Z_i)]\}.
\]

Since $e_k\sim\mathrm{Laplace}(0,b_n)$, for any interval of length $1$, based on the shape of the Laplace distribution, we have
\[
P(e_k\in[a,a+1])
\le
P(e_k\in[-\frac{1}{2},\frac{1}{2}])
=
1-e^{-\frac{1}{2b_n}}.
\]

Thus
\[
P(R_{ik}^c)
\le
1-e^{-\frac{1}{2b_n}}.
\]

i.e. $P(R_{ik})\ge e^{-\frac{1}{2b_n}}$. Then we obtain

\[
P(\bigcap_{k=1}^n R_{ik})\ge e^{-\frac{n}{2b_n}}
\]

So when $\frac{n}{b_n}\to0$ as $n\to\infty$, we have $P(\bigcap_{k=1}^n R_{ik})\to 1$. When $n$ is fixed and $b_n=b\to\infty$, we also have $\frac{n}{b}\to 0$ and derive $P(\bigcap_{k=1}^n R_{ik})\to 1$.

Therefore
$$
P(\bigcup_{i=1}^m\bigcap_{k=1}^n R_{ik})\ge P(\bigcap_{k=1}^n R_{ik})\to1, 
$$
which implies
\[
P(R)\to1.
\]

\subsection*{Proof of Theorem \ref{theorem2}}

First of all, we show the existence of a qualified $t_n$. Let $\{a_n\}$ be any sequence satisfying $a_n\to\infty$ and $\frac{a_n}{n}\to0$. Such $a_n$ could be found in any case (e.g. $a_n=\sqrt{n}$). Then let $t_n=-F_{n1}^{-1}(\frac{a_n}{n})$, where $F_{n1}^{-1}$ denotes the generalized inverse, i.e. $F_{n1}^{-1}(u)=\inf\{x\in\mathbb R:F_{n1}(x)\ge u\}$. According to this definition, we can see that $F^{-1}_{n1}$ exists for any distribution function $F_{n1}$, regardless of whether it is strictly increasing or not. Then we have $F_{n1}(-t_n)=\frac{a_n}{n}\to0$, and $nF_{n1}(-t_n)=a_n\to\infty$.

Furthermore, according to the proof of Theorem \ref{theorem3}, we have $F_{n1}$ uniformly converges to a fixed $ F_1$. Meanwhile, since the support of $F_1$ is the entire real line, we have $t_n=-F_{n1}^{-1}(\frac{a_n}{n})\to+\infty$. Therefore, the qualified $t_n$ always exists.

  Next, we show that $P(R)\to1$. Here we denote $E_n=\{\exists i_0 \in\{1,\dots,n\}, \text{ s.t. } f_n(Z_{i_0})<-t_n\}$. Then we have
  \begin{align*}
      P(R)\ge P(R|E_n)P(E_n)\ge P(\bigcap_{k=1}^n R_{i_0k})P(E_n)
  \end{align*}
  Here we know that $P(E_n)=1-[1-F_n(-t_n)]^n$. Since $F_n[-t_n]\to 0$, we know that $P(E_n)\sim 1-e^{-nF_n(-t_n)}\to 1$. So we have $P(E_n)\to 1$. Then for the first term, we have
  
  \begin{align*}
      P(\bigcap_{k=1}^n R_{i_0k})&=\prod_{k=1}^n P(R_{i_0k})=P(R_{i_01})^n=[1-P(R_{i_01})]^n\\
      &=\left[1-P(e_1\notin[f_n(X_{i_0})-1,f_n(X_{i_0})])\right]^n\\
      &\ge \left[1-P(e_1\notin[-t_n-1,-t_n])\right]^n\\
      &= \left[1-\frac{1}{2}e^{-\frac{t_n}{b_n}}(1-e^{-\frac{1}{b_n}})\right]^n
  \end{align*}
  Therefore, we have
  \begin{align*}
      \log(P(\bigcap_{k=1}^n R_{i_0k}))&\ge n\log\left[1-\frac{1}{2}e^{-\frac{t_n}{b_n}}(1-e^{-\frac{1}{b_n}})\right]\\
      &\sim -\frac{n}{2}e^{-\frac{t_n}{b_n}}(1-e^{-\frac{1}{b_n}})\\
      &\sim \frac{n}{2b_n}e^{-\frac{t_n}{b_n}}\to 0
  \end{align*}
  So we get $P(\bigcap_{k=1}^n R_{i_0k})\to 1$ as $n\to+\infty$. Because $P(R)\le 1$, we can then get that $P(R)\to 1$ as $n\to +\infty$.
  

\subsection*{Proof of Corollary \ref{corollary1}}
    
      Here for convenience we denote $F_L$ as the cdf of $\mathrm{Laplace}(0,b_n)$. Then it suffices to verify the conditions in Theorem \ref{theorem2}. We just let 
      $$
      t_n=b_n\log\frac{n}{\min(\log b_n,\log n)}
      $$
      It can be seen that, $t_n\ge b_n\log\frac{n}{\log n}\to +\infty$. Then we have $t_n\to +\infty$. Then we have
      \begin{align*}
          nF_L[-t_n]=\frac{1}{2}ne^{-\frac{t_n}{b_n}}=\frac{1}{2}\min(\log b_n,\log n)\to +\infty
      \end{align*}
      \begin{align*}
          F_L[-t_n]=\frac{1}{2}e^{-\frac{t_n}{b_n}}=\frac{1}{2n}\min(\log b_n,\log n)\le \frac{\log n}{2n}\to 0
      \end{align*}
      \begin{align*}
          \frac{n}{b_n}e^{-\frac{t_n}{b_n}}=\frac{n}{b_n}\frac{\min(\log b_n,\log n)}{n}\le\frac{\min(\log b_n,\log n)}{b_n}\to 0
      \end{align*}
      Therefore, all the conditions are verified, ans thus we can get $P(R)\to 1$ as $n\to +\infty$.

\subsection*{Proof of Corollary \ref{corollary2}}
  


  
      Here we can find that
      $$
      P(R)=P(\bigcup_{j=1}^p R_j)\ge P(R_{j_0})
      $$
      Based on this, we can directly apply Lemma \ref{theorem1} and Theorem \ref{theorem2} to $R_{j_0}$ to get the existence of $t_n$ and $P(R_{j_0})\to 1$, which completes the proof.

\clearpage
\section{Additional Experiment Results}
\begin{table}[htbp]
\centering
\caption{Correlation between watermarked and unwatermarked data under different settings. Values are the sample mean (standard deviation) based on 500 replications.}
\label{tab:distribution_results_transposed}
\begin{tabular}{c c c c}
\toprule
Distribution & $n=500$ & $n=2000$ & $n=5000$ \\
\midrule
$\mathrm{N}(0,1)$         & 0.963 (0.005) & 0.961 (0.003) & 0.960 (0.002) \\
$\mathrm{Exp}(1)$          & 0.878 (0.018) & 0.880 (0.010) & 0.882 (0.006) \\
$\mathrm{Unif}(0,1)$       & 0.895 (0.013) & 0.895 (0.007) & 0.895 (0.005) \\
$\mathrm{Beta}(2,5)$       & 0.932 (0.010) & 0.932 (0.006) & 0.931 (0.003) \\
$\mathrm{Laplace}(0,1)$    & 0.976 (0.004) & 0.979 (0.001) & 0.980 (0.001) \\
$\mathrm{Poisson}(1)$      & 0.948 (0.013) & 0.956 (0.005) & 0.958 (0.003) \\
$\mathrm{Bin}(5,0.3)$      & 0.895 (0.012) & 0.895 (0.006) & 0.895 (0.004) \\
$\mathrm{Bernoulli}(0.5)$  & 0.684 (0.016) & 0.684 (0.008) & 0.684 (0.005) \\
\bottomrule
\end{tabular}
\end{table}

\begin{table}[!ht]
\centering
\small
\setlength{\tabcolsep}{5.5pt}
\renewcommand{\arraystretch}{1.15}
\begin{threeparttable}
\caption{Robustness of watermark detection under subsampling for continuous distributions (25\% subset).}
\label{tab:table_robustness_continuous_prop1}
\begin{tabular}{@{}l l S[table-format=1.3] S[table-format=1.3]
S[table-format=1.3] S[table-format=1.3]
S[table-format=1.3] S[table-format=1.3]@{}}
\toprule
\multirow{2}{*}{Distribution} & \multirow{2}{*}{Method} &
\multicolumn{2}{c}{$n=500$} &
\multicolumn{2}{c}{$n=2000$} &
\multicolumn{2}{c}{$n=5000$} \\
\cmidrule(lr){3-4}\cmidrule(lr){5-6}\cmidrule(lr){7-8}
& & TP & TN & TP & TN & TP & TN \\
\midrule
$\mathrm{N}(0,1)$
& STAMP & 1.000 & 1.000 & 1.000 & 1.000 & 1.000 & 1.000 \\
& GLW & 1.000 & 0.962 & 1.000 & 0.960 & 1.000 & 0.952 \\
& TabularMark & 0.000 & 1.000 & 0.000 & 1.000 & 0.000 & 1.000 \\
\addlinespace

$\mathrm{Exp}(1)$
& STAMP & 1.000 & 1.000 & 1.000 & 1.000 & 1.000 & 1.000 \\
& GLW & 1.000 & 0.952 & 1.000 & 0.952 & 1.000 & 0.942 \\
& TabularMark & 0.000 & 1.000 & 0.000 & 1.000 & 0.000 & 1.000 \\
\addlinespace

$\mathrm{Unif}(0,1)$
& STAMP & 1.000 & 1.000 & 1.000 & 1.000 & 1.000 & 1.000 \\
& GLW & 1.000 & 0.954 & 1.000 & 0.946 & 1.000 & 0.958 \\
& TabularMark & 0.024 & 0.968 & 0.008 & 0.968 & 0.010 & 0.960 \\
\addlinespace

$\mathrm{Beta}(2,5)$
& STAMP & 1.000 & 1.000 & 1.000 & 1.000 & 1.000 & 1.000 \\
& GLW & 1.000 & 0.946 & 1.000 & 0.948 & 1.000 & 0.922 \\
& TabularMark & 0.026 & 0.964 & 0.024 & 0.980 & 0.026 & 0.960 \\
\addlinespace

$\mathrm{Laplace}(0,1)$
& STAMP & 1.000 & 1.000 & 1.000 & 1.000 & 1.000 & 1.000 \\
& GLW & 1.000 & 0.938 & 1.000 & 0.954 & 1.000 & 0.944 \\
& TabularMark & 0.000 & 1.000 & 0.000 & 1.000 & 0.000 & 1.000 \\
\bottomrule
\end{tabular}
\end{threeparttable}
\end{table}

\begin{table}[!ht]
\centering
\small
\setlength{\tabcolsep}{5.5pt}
\renewcommand{\arraystretch}{1.15}
\begin{threeparttable}
\caption{Robustness of watermark detection under subsampling for continuous distributions (50\% subset).}
\label{tab:robustness_pi_050}
\begin{tabular}{@{}l l S[table-format=1.3] S[table-format=1.3]
S[table-format=1.3] S[table-format=1.3]
S[table-format=1.3] S[table-format=1.3]@{}}
\toprule
\multirow{2}{*}{Distribution} & \multirow{2}{*}{Method} &
\multicolumn{2}{c}{$n=500$} &
\multicolumn{2}{c}{$n=2000$} &
\multicolumn{2}{c}{$n=5000$} \\
\cmidrule(lr){3-4}\cmidrule(lr){5-6}\cmidrule(lr){7-8}
& & TP & TN & TP & TN & TP & TN \\
\midrule
$\mathrm{N}(0,1)$
& STAMP & 1.000 & 1.000 & 1.000 & 1.000 & 1.000 & 1.000 \\
& GLW & 1.000 & 0.940 & 1.000 & 0.952 & 1.000 & 0.956 \\
& TabularMark & 0.000 & 1.000 & 0.000 & 1.000 & 0.000 & 1.000 \\

\addlinespace
$\mathrm{Exp}(1)$
& STAMP & 1.000 & 1.000 & 1.000 & 1.000 & 1.000 & 1.000 \\
& GLW & 1.000 & 0.946 & 1.000 & 0.950 & 1.000 & 0.966 \\
& TabularMark & 0.000 & 1.000 & 0.000 & 1.000 & 0.000 & 1.000 \\

\addlinespace
$\mathrm{Unif}(0,1)$
& STAMP & 1.000 & 1.000 & 1.000 & 1.000 & 1.000 & 1.000 \\
& GLW & 1.000 & 0.968 & 1.000 & 0.936 & 1.000 & 0.948 \\
& TabularMark & 0.024 & 0.968 & 0.008 & 0.968 & 0.010 & 0.960 \\

\addlinespace
$\mathrm{Beta}(2,5)$
& STAMP & 1.000 & 1.000 & 1.000 & 1.000 & 1.000 & 1.000 \\
& GLW & 1.000 & 0.958 & 1.000 & 0.950 & 1.000 & 0.944 \\
& TabularMark & 0.026 & 0.964 & 0.024 & 0.980 & 0.026 & 0.960 \\

\addlinespace
$\mathrm{Laplace}(0,1)$
& STAMP & 1.000 & 1.000 & 1.000 & 1.000 & 1.000 & 1.000 \\
& GLW & 1.000 & 0.944 & 1.000 & 0.948 & 1.000 & 0.928 \\
& TabularMark & 0.000 & 1.000 & 0.000 & 1.000 & 0.000 & 1.000 \\

\bottomrule
\end{tabular}
\end{threeparttable}
\end{table}

\begin{table}[!ht]
\centering
\small
\setlength{\tabcolsep}{5.5pt}
\renewcommand{\arraystretch}{1.15}
\begin{threeparttable}
\caption{Robustness of watermark detection under subsampling for continuous distributions (75\% subset).}
\label{tab:robustness_pi_075}
\begin{tabular}{@{}l l S[table-format=1.3] S[table-format=1.3]
S[table-format=1.3] S[table-format=1.3]
S[table-format=1.3] S[table-format=1.3]@{}}
\toprule
\multirow{2}{*}{Distribution} & \multirow{2}{*}{Method} &
\multicolumn{2}{c}{$n=500$} &
\multicolumn{2}{c}{$n=2000$} &
\multicolumn{2}{c}{$n=5000$} \\
\cmidrule(lr){3-4}\cmidrule(lr){5-6}\cmidrule(lr){7-8}
& & TP & TN & TP & TN & TP & TN \\
\midrule
$\mathrm{N}(0,1)$
& STAMP & 1.000 & 1.000 & 1.000 & 1.000 & 1.000 & 1.000 \\
& GLW & 1.000 & 0.934 & 1.000 & 0.952 & 1.000 & 0.942 \\
& TabularMark & 0.000 & 1.000 & 0.000 & 1.000 & 0.000 & 1.000 \\

\addlinespace
$\mathrm{Exp}(1)$
& STAMP & 1.000 & 1.000 & 1.000 & 1.000 & 1.000 & 1.000 \\
& GLW & 1.000 & 0.958 & 1.000 & 0.934 & 1.000 & 0.950 \\
& TabularMark & 0.000 & 1.000 & 0.000 & 1.000 & 0.000 & 1.000 \\

\addlinespace
$\mathrm{Unif}(0,1)$
& STAMP & 1.000 & 1.000 & 1.000 & 1.000 & 1.000 & 1.000 \\
& GLW & 1.000 & 0.938 & 1.000 & 0.942 & 1.000 & 0.944 \\
& TabularMark & 0.024 & 0.968 & 0.008 & 0.968 & 0.010 & 0.960 \\

\addlinespace
$\mathrm{Beta}(2,5)$
& STAMP & 1.000 & 1.000 & 1.000 & 1.000 & 1.000 & 1.000 \\
& GLW & 1.000 & 0.958 & 1.000 & 0.948 & 1.000 & 0.946 \\
& TabularMark & 0.026 & 0.964 & 0.024 & 0.980 & 0.026 & 0.960 \\

\addlinespace
$\mathrm{Laplace}(0,1)$
& STAMP & 1.000 & 1.000 & 1.000 & 1.000 & 1.000 & 1.000 \\
& GLW & 1.000 & 0.946 & 1.000 & 0.938 & 1.000 & 0.926 \\
& TabularMark & 0.000 & 1.000 & 0.000 & 1.000 & 0.000 & 1.000 \\

\bottomrule
\end{tabular}
\end{threeparttable}
\end{table}

\begin{table}[!ht]
\centering
\small
\setlength{\tabcolsep}{5.5pt}
\renewcommand{\arraystretch}{1.15}
\begin{threeparttable}
\caption{Robustness of watermark detection under subsampling for continuous distributions (10 observations selected).}
\label{tab:robustness_prop_10}
\begin{tabular}{@{}l l S[table-format=1.3] S[table-format=1.3]
S[table-format=1.3] S[table-format=1.3]
S[table-format=1.3] S[table-format=1.3]@{}}
\toprule
\multirow{2}{*}{Distribution} & \multirow{2}{*}{Method} &
\multicolumn{2}{c}{$n=500$} &
\multicolumn{2}{c}{$n=2000$} &
\multicolumn{2}{c}{$n=5000$} \\
\cmidrule(lr){3-4}\cmidrule(lr){5-6}\cmidrule(lr){7-8}
& & {TP} & {TN} & {TP} & {TN} & {TP} & {TN} \\
\midrule
$\mathcal{N}(0,1)$
& STAMP         & 1.000 & 1.000 & 1.000 & 1.000 & 1.000 & 1.000 \\
& GLW         & 1.000 & 0.976 & 1.000 & 0.988 & 1.000 & 0.984 \\
& TabularMark & 0.000 & 1.000 & 0.000 & 1.000 & 0.000 & 1.000 \\
\addlinespace[2pt]
$\mathrm{Exp}(1)$
& STAMP         & 1.000 & 1.000 & 1.000 & 1.000 & 1.000 & 1.000 \\
& GLW         & 1.000 & 0.986 & 1.000 & 0.984 & 1.000 & 0.974 \\
& TabularMark & 0.000 & 1.000 & 0.000 & 1.000 & 0.000 & 1.000 \\
\addlinespace[2pt]
$\mathrm{Unif}(0,1)$
& STAMP         & 1.000 & 1.000 & 1.000 & 1.000 & 1.000 & 1.000 \\
& GLW         & 1.000 & 0.964 & 1.000 & 0.982 & 1.000 & 0.974 \\
& TabularMark & 0.024 & 0.970 & 0.016 & 0.974 & 0.012 & 0.976 \\
\addlinespace[2pt]
$\mathrm{Beta}(2,5)$
& STAMP         & 1.000 & 1.000 & 1.000 & 1.000 & 1.000 & 1.000 \\
& GLW         & 1.000 & 0.974 & 1.000 & 0.984 & 1.000 & 0.982 \\
& TabularMark & 0.026 & 0.978 & 0.018 & 0.988 & 0.016 & 0.970 \\
\addlinespace[2pt]
$\mathrm{Laplace}(0,1)$
& STAMP         & 1.000 & 1.000 & 1.000 & 1.000 & 1.000 & 1.000 \\
& GLW         & 1.000 & 0.978 & 1.000 & 0.982 & 1.000 & 0.974 \\
& TabularMark & 0.000 & 1.000 & 0.000 & 1.000 & 0.000 & 1.000 \\
\bottomrule
\end{tabular}
\end{threeparttable}
\end{table}

\begin{table}[!ht]
\centering
\small
\setlength{\tabcolsep}{5.5pt}
\renewcommand{\arraystretch}{1.15}
\begin{threeparttable}
\caption{Robustness of watermark detection under subsampling for continuous distributions (100 observations selected).}
\label{tab:robustness_prop_100}
\begin{tabular}{@{}l l S[table-format=1.3] S[table-format=1.3]
S[table-format=1.3] S[table-format=1.3]
S[table-format=1.3] S[table-format=1.3]@{}}
\toprule
\multirow{2}{*}{Distribution} & \multirow{2}{*}{Method} &
\multicolumn{2}{c}{$n=500$} &
\multicolumn{2}{c}{$n=2000$} &
\multicolumn{2}{c}{$n=5000$} \\
\cmidrule(lr){3-4}\cmidrule(lr){5-6}\cmidrule(lr){7-8}
& & {TP} & {TN} & {TP} & {TN} & {TP} & {TN} \\
\midrule
$\mathrm{N}(0,1)$
& STAMP         & 1.000 & 1.000 & 1.000 & 1.000 & 1.000 & 1.000 \\
& GLW         & 1.000 & 0.960 & 1.000 & 0.952 & 1.000 & 0.952 \\
& TabularMark & 0.000 & 1.000 & 0.000 & 1.000 & 0.000 & 1.000 \\
\addlinespace[2pt]
$\mathrm{Exp}(1)$
& STAMP         & 1.000 & 1.000 & 1.000 & 1.000 & 1.000 & 1.000 \\
& GLW         & 1.000 & 0.950 & 1.000 & 0.946 & 1.000 & 0.960 \\
& TabularMark & 0.000 & 1.000 & 0.000 & 1.000 & 0.000 & 1.000 \\
\addlinespace[2pt]
$\mathrm{Unif}(0,1)$
& STAMP         & 1.000 & 1.000 & 1.000 & 1.000 & 1.000 & 1.000 \\
& GLW         & 1.000 & 0.944 & 1.000 & 0.954 & 1.000 & 0.952 \\
& TabularMark & 0.024 & 0.968 & 0.008 & 0.968 & 0.010 & 0.960 \\
\addlinespace[2pt]
$\mathrm{Beta}(2,5)$
& STAMP         & 1.000 & 1.000 & 1.000 & 1.000 & 1.000 & 1.000 \\
& GLW         & 1.000 & 0.944 & 1.000 & 0.948 & 1.000 & 0.962 \\
& TabularMark & 0.026 & 0.966 & 0.022 & 0.982 & 0.026 & 0.962 \\
\addlinespace[2pt]
$\mathrm{Laplace}(0,1)$
& STAMP         & 1.000 & 1.000 & 1.000 & 1.000 & 1.000 & 1.000 \\
& GLW         & 1.000 & 0.946 & 1.000 & 0.946 & 1.000 & 0.956 \\
& TabularMark & 0.000 & 1.000 & 0.000 & 1.000 & 0.000 & 1.000 \\
\bottomrule
\end{tabular}
\end{threeparttable}
\end{table}

\begin{table}[!ht]
  \centering
  \small
  \setlength{\tabcolsep}{6pt}
  \renewcommand{\arraystretch}{1.12}
  \begin{threeparttable}
  \caption{K–S distance across sample sizes for continuous and discrete distributions. Each entry reports the mean with standard deviation in parentheses.}
  \label{tab:table_ks_all}
  \begin{tabular}{@{} l l ccc @{}}
    \toprule
    \multirow{2}{*}{Distribution} & \multirow{2}{*}{Method} &
      \multicolumn{3}{c}{Sample Size $n$} \\
    \cmidrule(lr){3-5}
     & & 500 & 2000 & 5000 \\
    \midrule
    \multicolumn{5}{l}{\textbf{Continuous distributions}} \\
    \addlinespace[2pt]
    \multirow{3}{*}{$\mathrm{N}(0,1)$}
      & STAMP   & 0.0368 (0.0109) & 0.0189 (0.0057) & 0.0123 (0.0037) \\
      & GLW   & 0.0010 (0.0001) & 0.0020 (0.0003) & 0.0013 (0.0002) \\
      & TabularMark & 0.0076 (0.0009) & 0.0111 (0.0022) & 0.0085 (0.0013) \\
    \addlinespace[2pt]
    \multirow{3}{*}{$\mathrm{Exp}(1)$}
      & STAMP   & 0.0368 (0.0113) & 0.0185 (0.0054) & 0.0120 (0.0036) \\
      & GLW   & 0.0017 (0.0002) & 0.0028 (0.0004) & 0.0020 (0.0002) \\
      & TabularMark & 0.0277 (0.0014) & 0.0258 (0.0030) & 0.0270 (0.0019) \\
    \addlinespace[2pt]
    \multirow{3}{*}{$\mathrm{Unif}(0,1)$}
      & STAMP   & 0.0374 (0.0117) & 0.0186 (0.0056) & 0.0120 (0.0035) \\
      & GLW   & 0.0020 (0.0002) & 0.0033 (0.0004) & 0.0024 (0.0002) \\
      & TabularMark & 0.0386 (0.0012) & 0.0410 (0.0032) & 0.0389 (0.0018) \\
    \addlinespace[2pt]
    \multirow{3}{*}{$\mathrm{Beta}(2,5)$}
      & STAMP   & 0.0368 (0.0111) & 0.0192 (0.0061) & 0.0122 (0.0035) \\
      & GLW   & 0.0035 (0.0002) & 0.0050 (0.0006) & 0.0040 (0.0004) \\
      & TabularMark & 0.0549 (0.0018) & 0.0537 (0.0030) & 0.0540 (0.0023) \\
    \addlinespace[2pt]
    \multirow{3}{*}{$\mathrm{Laplace}(0,1)$}
      & STAMP   & 0.0374 (0.0117) & 0.0186 (0.0056) & 0.0120 (0.0035) \\
      & GLW   & 0.0011 (0.0001) & 0.0020 (0.0003) & 0.0014 (0.0002) \\
      & TabularMark & 0.0084 (0.0011) & 0.0118 (0.0023) & 0.0094 (0.0016) \\
    \addlinespace[4pt]
    \multicolumn{5}{l}{\textbf{Discrete distributions}} \\
    \addlinespace[2pt]
    \multirow{3}{*}{Pois(1)}
      & STAMP   & \textbf{0.0485} (0.0197) & \textbf{0.0486} (0.0105) & \textbf{0.0494} (0.0068) \\
      & GLW   & NA (NA) & NA (NA) & NA (NA) \\
      & TabularMark & NA (NA) & NA (NA) & NA (NA) \\
    \addlinespace[2pt]
    \multirow{3}{*}{Bin(5,0.3)}
      & STAMP   & \textbf{0.0242} (0.0113) & \textbf{0.0116} (0.0054) & \textbf{0.0077} (0.0036) \\
      & GLW   & NA (NA) & NA (NA) & NA (NA) \\
      & TabularMark & NA (NA) & NA (NA) & NA (NA) \\
    \addlinespace[2pt]
    \multirow{3}{*}{Ber(0.5)}
      & STAMP   & \textbf{0.0176} (0.0135) & \textbf{0.0086} (0.0065) & \textbf{0.0056} (0.0043) \\
      & GLW   & NA (NA) & NA (NA) & NA (NA) \\
      & TabularMark & NA (NA) & NA (NA) & NA (NA) \\
    \bottomrule
  \end{tabular}
  \begin{tablenotes}[flushleft]
    \footnotesize
    \item \textit{Notes:} For discrete distributions, K-S distances are not available for GLW and TabularMark and there results are shown as \texttt{NA (NA)}.
  \end{tablenotes}
  \end{threeparttable}
\end{table}

\begin{table}[htbp]
\centering
\footnotesize
\setlength{\tabcolsep}{6pt}
\renewcommand{\arraystretch}{1.15}
\begin{threeparttable}
\caption{Detection rates (TP and TN) in the logistic regression setting under different subset sizes.}
\label{tab:md_detection_subset}

\begin{tabular}{ll cc cc cc cc cc}
\toprule
& & \multicolumn{2}{c}{TAB-DRW} & \multicolumn{2}{c}{MUSE} & \multicolumn{2}{c}{TabWak} & \multicolumn{2}{c}{STAMP ($b=2.5$)} & \multicolumn{2}{c}{STAMP ($b=32$)} \\
\cmidrule(lr){3-4}\cmidrule(lr){5-6}\cmidrule(lr){7-8}\cmidrule(lr){9-10}\cmidrule(lr){11-12}
$(n,p)$ & Subset & TP & TN & TP & TN & TP & TN & TP & TN & TP & TN \\
\midrule

\multirow{3}{*}{$(500,\,3)$}
& 1   & NA & NA & 0.000 & 0.952 & 0.006 & 0.962 &1.000  & 0.258 & 1.000 & 0.996 \\
& 10  & 0.012 & 0.532 & 0.000 & 0.952 & 0.006 & 0.962 & 1.000 & 0.310 & 1.000 & 1.000 \\
& 100 & 0.858 & 0.532 & 1.000 & 0.952 & 0.104 & 0.962 & 1.000 & 0.688 & 1.000 & 1.000 \\
& Full & 0.850 & 0.532 & 1.000 & 0.952 & 0.182 & 0.962 & 1.000 & 0.994 & 1.000 & 1.000  \\
\addlinespace

\multirow{3}{*}{$(500,\,5)$}
& 1   & NA & NA & 0.000 & 0.952 & 0.098 & 0.924 & 1.000 & 0.408 & 1.000 & 1.000 \\
& 10  & 0.216 & 0.924 & 0.420 & 0.952 & 0.090 & 0.924 & 1.000 & 0.456 & 1.000 & 1.000 \\
& 100 & 0.258 & 0.924 & 0.986 & 0.952 & 0.092 & 0.924 & 1.000 & 0.376 & 1.000 & 1.000 \\
& Full & 0.172 & 0.924 & 1.000 & 0.952 & 0.136 & 0.924 & 1.000 & 0.994 & 1.000 & 1.000 \\
\addlinespace

\multirow{3}{*}{$(500,\,7)$}
& 1   & NA & NA & 0.000 & 0.964 & 0.056 & 0.944 & 1.000 & 0.594 & 1.000 & 1.000 \\
& 10  & 0.276 & 0.926 & 0.534 & 0.964 & 0.096 & 0.944 & 1.000 & 0.628 & 1.000 & 1.000 \\
& 100 & 0.464 & 0.926 & 0.998 & 0.964 & 0.102 & 0.944 & 1.000 & 0.824 & 1.000 & 1.000 \\
& Full & 0.368 & 0.926 & 1.000 & 0.964 & 0.246 & 0.944 & 1.000 & 0.996 & 1.000 & 1.000\\

\addlinespace
\multirow{3}{*}{$(1000,\,3)$}
& 1   & NA & NA & 0.000 & 0.944 & 0.010 & 0.970 & 1.000 &  0.046 & 1.000 & 0.998 \\
& 10  & 0.474 & 0.876 & 0.664 & 0.944 & 0.018 & 0.970 & 1.000 & 0.074 & 1.000 & 1.000 \\
& 100 & 0.316 & 0.876 & 0.998 & 0.944 & 0.034 & 0.970 & 1.000 & 0.386 & 1.000 & 1.000 \\
& Full & 0.668 & 0.876 & 1.000 & 0.944 & 0.016 & 0.970 & 1.000 & 0.998 & 1.000 & 1.000\\
\addlinespace

\multirow{3}{*}{$(1000,\,5)$}
& 1   & NA & NA & 0.448 & 0.946 & 0.048 & 0.944 & 1.000 & 0.092 & 1.000 & 1.000 \\
& 10  & 0.198 & 0.896 & 0.424 & 0.946 & 0.054 & 0.944 & 1.000 & 0.110 & 1.000 & 1.000 \\
& 100 & 0.354 & 0.896 & 0.944 & 0.946 & 0.112 & 0.944 & 1.000 & 0.398 & 1.000 & 1.000 \\
& Full & 0.002 & 0.896 & 1.000 & 0.946 & 0.446 & 0.944 & 1.000 & 0.992 & 1.000 & 1.000\\
\addlinespace

\multirow{3}{*}{$(1000,\,7)$}
& 1   & NA & NA & 0.000 & 0.952 & 0.092 & 0.954 & 1.000 & 0.110 & 1.000 & 1.000 \\
& 10  & 0.104 & 0.876 & 0.690 & 0.952 & 0.046 & 0.954 & 1.000 & 0.152 & 1.000 & 1.000 \\
& 100 & 0.248 & 0.876 & 0.996 & 0.952 & 0.088 & 0.954 & 1.000 & 0.406 & 1.000 & 1.000 \\
& Full & 0.250 & 0.876 & 1.000 & 0.952 & 0.242 & 0.954 & 1.000 & 0.996 & 1.000 & 1.000\\

\bottomrule
\end{tabular}
\begin{tablenotes}[flushleft]
\footnotesize
\item \textit{Notes:} Values are detection rates computed from 500 replications. TP and TN denote the true positive rate and true negative rate, respectively. Subset size indicates the number of observations used for detection; ``Full'' corresponds to using the entire training set. STAMP ($b=2.5$) and STAMP ($b=32$) denote the settings where the scale parameter of the Laplace noise is set to 2.5 and 32, respectively.
\end{tablenotes}
\end{threeparttable}
\end{table}

\begin{table}[htbp]
\centering
\caption{MSE results in the logistic regression setting.}
\label{tab:mse_results}

\begin{tabular}{l
S[table-format=1.4]
S[table-format=1.4]
S[table-format=1.4]
S[table-format=1.4]}
\toprule
Dataset & {TAB-DRW} & {MUSE} & {TabWak} & {STAMP} \\
\midrule
$n = 500,\, p = 3$   & {18.002} & {0.473} & {2.102} & {0.077} \\
$n = 500,\, p = 5$   & {1.521} & {1.696} & {1.738} & {0.085} \\
$n = 500,\, p = 7$   & {2.281} & {1.942} & {1.374} & {0.097} \\
$n = 1000,\, p = 3$  & {3.489} & {0.344} & {0.556} & {0.039} \\
$n = 1000,\, p = 5$  & {10.860} & {6.239} & {5.433} & {0.044} \\
$n = 1000,\, p = 7$  & {2.172} & {1.707} & {1.817} & {0.047} \\
\bottomrule
\end{tabular}

\end{table}

\begin{table}[htbp]
\centering
\small
\setlength{\tabcolsep}{6pt}
\caption{K--L divergence results in the logistic regression setting.}
\label{tab:kl_logistic_results}
\begin{tabular}{lcccc}
\toprule
Dataset & TAB-DRW & Muse & TabWak & STAMP \\
\midrule
$n=500, p=3$   & NA & NA & NA & 0.114 \\
$n=500, p=5$   & NA & NA & NA & 0.212 \\
$n=500, p=7$   & NA & NA & NA & 0.296 \\
$n=1000, p=3$  & NA & NA & NA & 0.110 \\
$n=1000, p=5$  & NA & NA & NA & 0.205 \\
$n=1000, p=7$  & NA & NA & NA & 0.285 \\
\bottomrule
\end{tabular}
\end{table}


\begin{table}[htbp]
\centering
\small
\setlength{\tabcolsep}{7pt}
\begin{threeparttable}
\caption{Prediction performance of XGBoost on the NHANES Diabetes dataset.}
\label{tab:nhanes_diabetes_xgb}

\begin{tabular}{l
                S[table-format=1.3]
                S[table-format=1.3]
                S[table-format=1.3]
                S[table-format=1.3]
                S[table-format=1.3]}
\toprule
Method & {AUC} & {Accuracy} & {Precision} & {Recall} & {F1 Score} \\
\midrule
TAB-DRW  & {0.626(0.112)} & {0.493(0.210)} & {0.044(0.042)} & {0.504(0.288)} & {0.081(0.060)} \\
MUSE     & {0.629(0.094)} & {0.534(0.205)} & {0.048(0.049)} & {0.482(0.292)} & {0.086(0.067)} \\
TabWak   & {0.635(0.101)} & {0.520(0.209)} & {0.047(0.043)} & {0.526(0.284)} & {0.087(0.062)} \\
STAMP      & {0.773(0.063)} & {0.747(0.107)} & {0.120(0.051)} & {0.763(0.138)} & {0.201(0.066)} \\
Original & {0.809(0.055)} & {0.791(0.079)} & {0.141(0.053)} & {0.787(0.112)} & {0.234(0.070)} \\
\bottomrule
\end{tabular}

\begin{tablenotes}[flushleft]
\footnotesize
\item \textit{Notes:} Values are reported as mean (standard deviation) over 500 replications. All metrics lie in $[0,1]$. Models are trained on watermarked data (by each method) and evaluated on the original test data.
\end{tablenotes}
\end{threeparttable}
\end{table}

\clearpage
\bibliographystyle{plain}  
\bibliography{references}  

@article{fang2025muse,
  title = {{{MUSE}}: Model-agnostic Tabular Watermarking via Multi-Sample Selection},
  author = {Fang, Liancheng and Liu, Aiwei and Zou, Henry Peng and Chen, Yankai and Zhang, Hengrui and Deng, Zhongfen and Yu, Philip S},
  year = 2025,
  journal = {arXiv:2505.24267},
  eprint = {2505.24267},
  archiveprefix = {arXiv}
}

@inproceedings{he2023algorithmically,
  title={Algorithmically effective differentially private synthetic data},
  author={He, Yiyun and Vershynin, Roman and Zhu, Yizhe},
  booktitle={Annual Conference on Learning Theory},
  pages={3941--3968},
  year={2023}
}

@article{xian2024raw,
  title={Raw: A robust and agile plug-and-play watermark framework for ai-generated images with provable guarantees},
  author={Xian, Xun and Wang, Ganghua and Bi, Xuan and Srinivasa, Jayanth and Kundu, Ashish and Hong, Mingyi and Ding, Jie},
  journal={Advances in Neural Information Processing Systems},
  volume={37},
  pages={132077--132105},
  year={2024}
}

@article{he2024watermarking,
  title = {Watermarking Generative Tabular Data},
  author = {He, Hengzhi and Yu, Peiyu and Ren, Junpeng and Wu, Ying Nian and Cheng, Guang},
  year = 2024,
  journal = {arXiv:2405.14018},
  eprint = {2405.14018},
  archiveprefix = {arXiv}
}

@article{ngo2024adaptive,
  title = {Adaptive and Robust Watermark for Generative Tabular Data},
  author = {Ngo, Dung Daniel and Ray, Archan and Seshadri, Akshay and Scott, Daniel and Obitayo, Saheed and Kumar, Niraj and Potluru, Vamsi K and Pistoia, Marco and Veloso, Manuela},
  year = 2024,
  journal = {arXiv:2409.14700},
  eprint = {2409.14700},
  archiveprefix = {arXiv}
}

@article{zhao2025tab,
  title = {{{TAB-DRW}}: A {{DFT-based}} Robust Watermark for Generative Tabular Data},
  author = {Zhao, Yizhou and Li, Xiang and Song, Peter and Long, Qi and Su, Weijie},
  year = 2025,
  journal = {arXiv:2511.21600},
  eprint = {2511.21600},
  archiveprefix = {arXiv}
}

@inproceedings{zheng2024tabularmark,
  title = {Tabularmark: {{Watermarking}} Tabular Datasets for Machine Learning},
  booktitle = {Proceedings of the 2024 on {{ACM SIGSAC}} Conference on Computer and Communications Security},
  author = {Zheng, Yihao and Xia, Haocheng and Pang, Junyuan and Liu, Jinfei and Ren, Kui and Chu, Lingyang and Cao, Yang and Xiong, Li},
  year = 2024,
  pages = {3570--3584}
}

@inproceedings{zhu2025tabwak,
  title = {Tabwak: {{A}} Watermark for Tabular Diffusion Models},
  booktitle = {International Conference on Learning Representations},
  author = {Zhu, Chaoyi and Tang, Jiayi and Galjaard, Jeroen M and Chen, Pin-Yu and Birke, Robert and Bos, Cornelis and Chen, Lydia Y and others},
  year = 2025,
  pages = {1--28},
  publisher = {OpenReview. net}
}

@article{zizhuo2024robust,
  title={Robust blind image watermarking based on interest points},
  author={Wang, Zizhuo and Hu, Kun and Hu, Zixuan and Yang, Shuo and Wang, Xingjun and others},
  journal={Virtual Reality \& Intelligent Hardware},
  volume={6},
  number={4},
  pages={308--322},
  year={2024},
  publisher={Elsevier}
}

@article{tian2024perceptive,
  title={Perceptive self-supervised learning network for noisy image watermark removal},
  author={Tian, Chunwei and Zheng, Menghua and Li, Bo and Zhang, Yanning and Zhang, Shichao and Zhang, David},
  journal={IEEE Transactions on Circuits and Systems for Video Technology},
  volume={34},
  number={8},
  pages={7069--7079},
  year={2024},
  publisher={IEEE}
}

@inproceedings{muller2025black,
  title={Black-box forgery attacks on semantic watermarks for diffusion models},
  author={M{\"u}ller, Andreas and Lukovnikov, Denis and Thietke, Jonas and Fischer, Asja and Quiring, Erwin},
  booktitle={Proceedings of the Computer Vision and Pattern Recognition Conference},
  pages={20937--20946},
  year={2025}
}

@article{por2012unispach,
  title={UniSpaCh: A text-based data hiding method using Unicode space characters},
  author={Por, Lip Yee and Wong, KokSheik and Chee, Kok Onn},
  journal={Journal of Systems and Software},
  volume={85},
  number={5},
  pages={1075--1082},
  year={2012},
  publisher={Elsevier}
}

@article{yang2023watermarking,
  title={Watermarking text generated by black-box language models},
  author={Yang, Xi and Chen, Kejiang and Zhang, Weiming and Liu, Chang and Qi, Yuang and Zhang, Jie and Fang, Han and Yu, Nenghai},
  journal={arXiv:2305.08883},
  year={2023}
}

@inproceedings{atallah2001natural,
  title={Natural language watermarking: Design, analysis, and a proof-of-concept implementation},
  author={Atallah, Mikhail J and Raskin, Victor and Crogan, Michael and Hempelmann, Christian and Kerschbaum, Florian and Mohamed, Dina and Naik, Sanket},
  booktitle={International Workshop on Information Hiding},
  pages={185--200},
  year={2001},
  organization={Springer}
}

@article{bi2023distribution,
  title={Distribution-invariant differential privacy},
  author={Bi, Xuan and Shen, Xiaotong},
  journal={Journal of Econometrics},
  volume={235},
  number={2},
  pages={444--453},
  year={2023},
  publisher={Elsevier}
}

@InProceedings{Zhu_2018_ECCV,
author = {Zhu, Jiren and Kaplan, Russell and Johnson, Justin and Fei-Fei, Li},
title = {HiDDeN: Hiding Data with Deep Networks},
booktitle = {Proceedings of the European Conference on Computer Vision (ECCV)},
month = {September},
year = {2018}
}

@inproceedings{Tancik_2020_CVPR,
author = {Tancik, Matthew and Mildenhall, Ben and Ng, Ren},
title = {StegaStamp: Invisible Hyperlinks in Physical Photographs},
booktitle = {Proceedings of the IEEE/CVF Conference on Computer Vision and Pattern Recognition (CVPR)},
month = {June},
year = {2020}
}

@article{altun2009optimal,
  title={Optimal spread spectrum watermark embedding via a multistep feasibility formulation},
  author={Altun, H Oktay and Orsdemir, Adem and Sharma, Gaurav and Bocko, Mark F},
  journal={IEEE Transactions on Image Processing},
  volume={18},
  number={2},
  pages={371--387},
  year={2009},
  publisher={IEEE}
}

@article{hernandez2000dct,
  title={DCT-domain watermarking techniques for still images: Detector performance analysis and a new structure},
  author={Hernandez, Juan R and Amado, Martin and Perez-Gonzalez, Fernando},
  journal={IEEE Transactions on Image Processing},
  volume={9},
  number={1},
  pages={55--68},
  year={2000},
  publisher={IEEE}
}

@inproceedings{gunn2025undetectable,
  title={An undetectable watermark for generative image models},
  author={Gunn, Samuel and Zhao, Xuandong and Song, Dawn},
  booktitle={International Conference on Learning Representations},
  volume={2025},
  pages={6612--6637},
  year={2025}
}

@inproceedings{hu2024unbiased,
  title={Unbiased watermark for large language models},
  author={Hu, Zhengmian and Chen, Lichang and Wu, Xidong and Wu, Yihan and Zhang, Hongyang and Huang, Heng},
  booktitle={International Conference on Learning Representations},
  volume={2024},
  pages={45408--45436},
  year={2024}
}

@misc{WatermarkingLargeLanguage2023,
  title = {Watermarking of Large Language Models},
  author = {Aaronson, Scott},
  year = 2023,
  month = aug,
  urldate = {2026-05-13},
  howpublished = {https://simons.berkeley.edu/talks/scott-aaronson-ut-austin-openai-2023-08-17},
  langid = {english}
}

@article{li2025statistical,
  title={A statistical framework of watermarks for large language models: Pivot, detection efficiency and optimal rules},
  author={Li, Xiang and Ruan, Feng and Wang, Huiyuan and Long, Qi and Su, Weijie J},
  journal={The Annals of Statistics},
  volume={53},
  number={1},
  pages={322--351},
  year={2025},
  publisher={Institute of Mathematical Statistics}
}

@inproceedings{lu2025wasa,
  title={WASA: WAtermark-based Source Attribution for Large Language Model-Generated Data},
  author={Lu, Xinyang and Wang, Jingtan and Zhao, Zitong and Dai, Zhongxiang and Foo, Chuan-Sheng and Ng, See Kiong and Low, Bryan Kian Hsiang},
  booktitle={Findings of the Association for Computational Linguistics: ACL 2025},
  pages={23791--23824},
  year={2025}
}

@article{zhang2024personamark,
  title={Personamark: Personalized llm watermarking for model protection and user attribution},
  author={Zhang, Yuehan and Lv, Peizhuo and Liu, Yinpeng and Ma, Yongqiang and Lu, Wei and Wang, Xiaofeng and Liu, Xiaozhong and Liu, Jiawei},
  journal={arXiv:2409.09739},
  year={2024}
}

@inproceedings{10.1007/11787006_1,
  title = {Differential Privacy},
  booktitle = {Automata, Languages and Programming},
  author = {Dwork, Cynthia},
  editor = {Bugliesi, Michele and Preneel, Bart and Sassone, Vladimiro and Wegener, Ingo},
  year = 2006,
  pages = {1--12},
  publisher = {Springer Berlin Heidelberg},
  address = {Berlin, Heidelberg},
  isbn = {978-3-540-35908-1}
}

@article{MORO201422,
  title = {A Data-Driven Approach to Predict the Success of Bank Telemarketing},
  author = {Moro, S{\'e}rgio and Cortez, Paulo and Rita, Paulo},
  year = 2014,
  journal = {Decision Support Systems},
  volume = {62},
  pages = {22--31},
  issn = {0167-9236},
  doi = {10.1016/j.dss.2014.03.001}
}

@inproceedings{kirchenbauer2023watermark,
  title={A watermark for large language models},
  author={Kirchenbauer, John and Geiping, Jonas and Wen, Yuxin and Katz, Jonathan and Miers, Ian and Goldstein, Tom},
  booktitle={International Conference on Machine Learning},
  pages={17061--17084},
  year={2023},
  organization={PMLR}
}






\end{document}